\documentclass[prb,
twocolumn,
superscriptaddress,showpacs,amsmath,amssymb]{revtex4}

\usepackage{hyperref}
\usepackage{graphicx}
\graphicspath{{./Kuvaajat/}}

\usepackage{color}

\newcommand{\tr}{\textrm{tr}}
\newcommand{\abs}[1]{\left \vert #1\right \vert}
\newcommand{\im}{i}
\newcommand{\unitvec}[1]{\hat{\vek{#1}}}
\newcommand{\vek}[1]{\mathbf{#1}}
\newcommand{\doo}{\partial}
\newcommand{\hc}{\textrm{ h.c.}}

\renewcommand{\l}{\hat{\mathbf{l}}}
\newcommand{\m}{\hat{\mathbf{m}}}
\newcommand{\n}{\hat{\mathbf{n}}}
\newcommand{\x}{\hat{\mathbf{x}}}
\newcommand{\y}{\hat{\mathbf{y}}}
\newcommand{\z}{\hat{\mathbf{z}}}

\newcommand{\khat}{\hat{p}}

\newcommand{\eps}{\varepsilon_\mathbf{p}}

\DeclareMathOperator{\levic}{\varepsilon^{\mu\nu\rho\sigma}}

\newcommand{\pf}{p_F}

\newcommand{\half}{\frac{1}{2}}
\renewcommand{\part}{\partial_t}

\newcommand{\mom}{\hat{p}}
\newcommand{\gap}{\Delta}
\newcommand{\gapc}{\Delta^\dagger}
\newcommand{\ch}{\chi} 
\newcommand{\tb}{T_{B}^3}
\newcommand{\te}{T_{E}^3}
\newcommand{\sgn}{\text{sgn}}

\begin{document}

\title{Torsional Landau levels and geometric anomalies in condensed matter Weyl systems}

\author{Sara Laurila}
\email{sara.laurila@helsinki.fi}

\author{Jaakko Nissinen}
\email{jaakko.nissinen@aalto.fi}

\affiliation{Low Temperature Laboratory, Department of Applied Physics, Aalto University, P.O. Box 15100, FI-00076
Aalto, Finland}

\date{\today}

\begin{abstract}
We consider the role of coordinate dependent tetrads (``Fermi velocities"), momentum space geometry, and torsional Landau levels (LLs) in condensed matter systems with low-energy Weyl quasiparticles. In contrast to their relativistic counterparts, they arise at finite momenta and an explicit cutoff to the linear spectrum. Via the universal coupling of tetrads to momentum, they experience geometric chiral and axial anomalies with gravitational character. {More precisely, at low-energy, the fermions experience} background fields corresponding to emergent anisotropic Riemann-Cartan and Newton-Cartan spacetimes, depending on the form of the low-energy dispersion.  On these backgrounds, we show how torsion and the Nieh-Yan (NY) anomaly appear {in condensed matter Weyl systems} with a ultraviolet (UV) parameter with dimensions of momentum. The torsional NY anomaly arises in simplest terms from the spectral flow of torsional LLs coupled to the nodes at finite momenta and the linear approximation with a cutoff. We carefully review {the torsional anomaly and spectral flow} for relativistic fermions at zero momentum {and} contrast this with the spectral flow, non-zero torsional anomaly and the appearance the dimensionful UV-cutoff parameter in condensed matter systems at finite momentum. We apply this to chiral transport anomalies sensitive to the emergent tetrads in non-homogenous chiral superconductors, superfluids and Weyl semimetals under elastic strain. {This leads to the suppression of anomalous density at nodes from geometry, as compared to (pseudo)gauge fields}. We also briefly discuss the role torsion in anomalous thermal transport for non-relativistic Weyl fermions, which arises via Luttinger's fictitious gravitational field corresponding to thermal gradients.
\end{abstract}

\maketitle

\section{Introduction}

Gapless {fermionic} quasiparticles with linear spectrum protected by topology arise in many {condensed matter systems} in three dimensions \cite{NielsenNinomiya83, CallanHarvey85, Volovik03, Horava05, WanEtAl11}. In particular, accidental crossings of two inversion ($P$) or time-reversal ($T$) breaking bands at the Fermi energy lead to {stable} quasirelativistic particles with low-energy dispersion analogous to relativistic Weyl fermions \cite{Herring37, AbrisokovBenelavskii71}. Fourfold degenerate crossings with Dirac-like low-energy excitations occur for combined $P,T$ (and/or other similar protecting) symmetries \cite{BalatskyEtAl14, ArmitageEtAl18}. Similarly, in chiral superconductors and superfluids with gap nodes, Majorana-Weyl excitations arise at low energy \cite{Volovik85, Volovik1986a, Volovik1986b, Volovik90, Volovik03, ReadGreen00}.

By a very general theorem from topology \cite{Horava05}, the low-energy linear theory near the {three-dimensional} Fermi point node takes universally the ($\gamma$-matrix) form of a quasirelativistic Weyl/Dirac spectrum, with the precise form of the metric and other background fields depending on the microscopic details. It is then of interest to study the detailed form of this emergent Dirac operator with an explicit cutoff and compare to fundamental, Lorentz invariant fermions. Following this logic, the concept of so-called {momentum space} pseudo gauge fields \cite{Volovik03, ShapourianEtAl15, CortijoEtAl15, Landsteiner16, Fujimoto16, PikulinEtAl16, GrushinEtAl16, SukhachovEtAl17, SukhachovEtAl18, SukhachovEtAl18b, IlanEtAl19} and ``emergent" spacetime \cite{Volovik1986a, Volovik1986b, Volovik90, Volovik03, ReadGreen00, MesarosEtAl10, Son13, BanerjeeEtAl14, NissinenVolovik17, WeststromOjanen17, NissinenVolovik2018, GolanStern18, Nissinen2019, LiangOjanen19, WilsonEtAl20, JafariEtAl20} in non-relativistic condensed matter systems has emerged, where the low-energy fermions can experience background fields of various physical origins, similar to what appears for spin-1/2 (or even higher spin) fermions on curved spacetimes in general relativity or its non-relativistic generalizations with non-relativistic coordinate invariance. 

Notably, {in the low-energy quasilinear theory, the local Fermi velocities form emergent tetrads which determine the geometry of the conical dispersion}. {The tetrads, and its field strength torsion}, couple to the quasiparticle momentum {effectively} as in gravity. {The effects of such fields in non-relativistic systems appearing at finite density $\mu_F$ and Fermi-momentum $p_F$ are expected to be very different from their relativistic counterparts appearing at $p=0$}. {Amongst other things}, the system at finite Fermi or crystal momentum is then charged under the field strength these geometric background fields \cite{MesarosEtAl10, JuricicEtAl12, ParrikarEtAl14, ShapourianEtAl15, PachosEtAl20}. In three spatial dimensions, this corresponds to the anomalous translational symmetry for chiral fermions, leading to axial anomalies in the system \cite{Nissinen2019, Burkov20} from momentum space translations. For other relevant condensed matter considerations of this anomaly, see e.g. \cite{VolovikMineev81, Volovik84, Volovik1986a, CombescotDombre86, BalatskiiEtAl86, Volovik87, Volovik95, BevanEtAl97, Volovik03, ZyuzinBurkov12, SonYamamoto12, Zahed12, ZhouEtAl13, SonSpivak13, BasarKharzeevZahed2013, Landsteiner2014, LucasEtAl2016, GoothEtAl17, ArmitageEtAl18}. {In this paper we point out that geometric (gravitational) contributions in the chiral anomaly, second order in gradients, are expected in generic non-homogenous condensed matter Weyl systems with momentum space fields (background spacetimes) due to inhomogenous deformations leading to torsion.}

{More generally, the appereance of the tetrad background fields in condensed matter Weyl systems is built-in in the low-energy theory}, thus opening the possibility of simulating Riemann-Cartan (or Newton-Cartan) spacetimes for the low-energy fermions. In case of non-trivial background torsion, the so-called chiral gravitational Nieh-Yan anomaly can appear \cite{NiehYan82, NiehYan82b}. In contrast to the axial anomaly with gauge fields, this anomaly depends on a non-universal UV cut-off parameter $\Lambda$, with canonical dimensions of momentum. While the status of the torsional contribution in relativistic systems has been debated for long \cite{Yajima96, ChandiaZanelli97, ObukhovEtAl97, Soo99, PeetersWaldron99, Comments, KubotaEtAl01}, the appearance of this term in non-relativistic condensed matter systems with explicit UV cutoff to the Weyl physics is a priori plausible \cite{ParrikarEtAl14, Nissinen2019}. {Aspects of the gravitational anomaly in condensed matter have} been considered in e.g. \cite{Zahed12, ZhouEtAl13, SunWan14, ParrikarEtAl14, PalumboPachos16, MaranerPachosPalumbo18, FerreirosEtAl19, CopettiLandsteiner19, Nissinen2019, Copetti20, Stone2019b} {including Weyl/Dirac fermions in superfluids, superconductors and semimetals. The dimensional hierarchy and descent relations of the torsional anomaly were recently analyzed in Ref. \cite{Stone2019b} from a Hamiltonian persperctive in a relativistic model}. {Nevertheless, it seems that any explicit value of the cutoff parameter has not been discussed in detail, except in the recent paper \cite{Nissinen2019} by one of the present authors.}  {In the simplest possible terms}, the non-universal anomaly UV scale originates from the regime of validity of the quasirelativistic linear spectrum {and the associated anomalous transport}. {This UV-scale is, in the simplest possible terms, just the validity of the Taylor expansion close to the node which is experimentally a low-energy scale in the system \cite{Nissinen2019}}. {Generalizing this}, it seems that the NY anomaly non-universally probes the chiral spectrum and transport, well-defined only at low energies, and conversely, merging in some left-right asymmetric way to other bands as required by global consistency and symmetries. {Indeed, at face value, the spectrum and spectral flow can be terminated in a multitude of inequivalent ways}. If the system is anisotropic, the interplay of different scales in the system becomes essential, as evidenced by {the consideration of the anomaly} in e.g. Newton-Cartan geometry with quadratic spectrum along a preferred direction or finite temperature (see below). 

Here we will further argue for {the torsional anomaly} term using the simplest {computational} apparatus for the chiral and axial anomaly: adiabatic spectral flow in the presence of torsional Landau levels \cite{NielsenNinomiya83, Volovik85}. In this context, the torsional LLs appeared implicitly already in Refs. \cite{Volovik85, BalatskiiEtAl86} and more recently in topological semimetals in \cite{ParrikarEtAl14} in {comparison with} Pauli-Villars regularization of Lorentz invariant fermions. On the other hand, such a relativistic regularization scheme is at best only an approximation in condensed matter systems, since the linear Weyl regime applies to low-energies with an explicit cutoff scale. This linear regime can be anisotropic and, furthermore, is continuously connected with the non-relativistic regime with quadratic dispersion. {Moreover, as discussed in this paper, the role of the spectral flow is drastically altered by the finite node momentum as compared to relativistic fermions.}

{The role of momentum space pseudo gauge fields, with momentum dependent axial charge also becomes evident in the geometric framework for the axial anomaly. Importantly, it is incorrect to assume the universal U(1) axial anomaly for such gauge fields, since the effective momentum space description has a finite regime of validity. To the best of our knowledge, it seems that this fact has been overlooked thus far.} {Related to the momentum dependence in the anomaly, the UV-scale can be supplemented by a infrared (IR) temperature scale of thermal fluctuations, in contrast to, say U(1), gauge fields}. {With some caveats, this IR anomaly becomes universal due to universality of thermal fluctuations close to the node. The thermal torsional anomaly and the associated currents were recently considered in Ref. \cite{NissinenVolovik2019}}. Contribution to the torsional NY anomaly at finite temperatures was further discussed in \cite{ImakiYamamoto19, Stone2019, NissinenVolovik19b, LiangOjanen19b, Imaki20} for relativistic fermions at $p=0$. The closely related role of torsion and viscoelastic thermal transport has been also studied e.g. in \cite{Shitade14, BradlynRead15, GromovAbanov15, Sekine16}. Here we will mostly focus on the non-universal UV contribution at zero temperature. For completeness, we comment on thermal effects by non-zero temperature gradients, which point to still new types of {anisotropic} torsional anomalies terms not present in systems with Lorentz invariance.
 
This rest of this paper is organized as follows. Section \ref{sec:spacetimes} discusses the low-energy Weyl Hamiltonian and the associated geometry in condensed matter systems from the perspective of emergent background spacetimes. The following Section \ref{sec:torsional_LLs} reviews the relativistic torsional anomaly and spectral flow argument, focusing on the extension to finite node momentum and the comparison with the anomaly for U(1) gauge fields presented in Appendix \ref{sec:appendix_EM}. Sec. \ref{sec:chiral} discusses the torsional anomaly in chiral superfluids and superconductors, where it can be matched with experiment \cite{Volovik95, BevanEtAl97, Nissinen2019}. This followed by a model of $T$-breaking strained semimetals in Sec. \ref{sec:WSM}. We also briefly discuss the role of torsion in the presence of thermal gradients in Sec \ref{sec:thermal}. We conclude with a comparison on previous results Sec. \ref{sec:comparison} and the conclusions and outlook of our results.

\section{Weyl fermions in condensed matter and relativistic systems}\label{sec:spacetimes}
 
\subsection{Weyl fermions in condensed matter}
We consider a {fermionic} system with broken time-reversal symmetry ($T$) or inversion ($P$). {In the vicinity of a generic degenerate crossing at $\vek{p}_W$, ignoring all other bands, the $2\times2$ Hamiltonian is $H = \sigma^a H_a$ in terms of the unit and Pauli matrices $\sigma^a$, $a=0,1,2,3$. This leads to the expansion}
\begin{align}
H(\vek{p}) = \sigma^a e_a^{i}(p-p_W)_{i} + \cdots \label{eq:HWeyl},
\end{align}
where
\begin{align}
e_a^i = \frac{\doo H_a}{\doo p_i}\bigg \vert_{p=p_W}. \label{eq:Taylor_tetrad}
\end{align}
The expansion is, of course, valid for $\abs{\vek{p}-\vek{p}_W}\ll p_W$ since the remainder is of the order of $\abs{\vek{p}-\vek{p}_W}^2$. This provides an explicit cutoff for the linear Weyl regime that is, nevertheless, continuously connected with the non-relativistic quadratic dispersing spectrum and the other bands. 

{The existence of the Weyl node degeneracy is protected by topologuy in a finite region} since there are three parameters and three constraints \cite{Herring37, AbrisokovBenelavskii71, Volovik03, Horava05}. Via rotations and scalings, $\tilde{p}_a = e^i_a p_i$, the Hamiltonian becomes the right- or left-handed relativistic Weyl Hamiltonian, at Fermi momentum $\tilde{p}_W$,
\begin{align}
\tilde{H}(\vek{p}) = \chi \sigma^a (\tilde{p}-\tilde{p}_W)_a
\end{align}
where $\chi=\pm 1 = \sgn(\det e^i_a)$ is the chirality, defined as the direction of (pseudo)spin with respect to the propagation momentum. The band energies are $E =(\tilde{p}-\tilde{p}_W)_0\pm \sqrt{\abs{\tilde{\vek{p}}-\tilde{\vek{p}}_W}^2}$. The role of the coefficients $e^\mu_a$ is simply to determine the (anisotropic) Fermi velocities of the {conical} dispersion $\omega^2 = -g^{ij}(p-p_W)_i (p-p_W)_j$ via the (inverse) metric
\begin{align}
g^{ij} = -\sum_{a,b=0,1,2,3} e^i_a e^j_b \delta^{ab} \equiv -e^i_a e^j_b \delta^{ab}  \label{eq:cone_metric}
\end{align}
where the Einstein summation convention for repeated latin and greek indices will be henceforth assumed. The spatial tetrad $e_a^i$ is extended to a non-degenerate matrix $e^{\mu}_a$ by considering the operator $\sigma^a e_a^{\mu}\im\doo_{\mu} =\im \doo_t - H(\vek{p})$ with $\mu=t,x,y,z$. In particular, the coefficient $e^\mu_0=\{1,v^i\}$ is non-trivial in type-II Weyl semimetals and {in} superfluids and superconductors with superflow. The case with non-zero spatial $e^{t}_a$, $a=1,2,3$ was considered in \cite{NissinenVolovik17}. These break different symmetries, while the spacelike tetrads {transform} like gauge potentials corresponding to axial magnetic and electric fields.  While the Hamiltonian \eqref{eq:HWeyl} is usually analyzed for translationally invariant systems, it remains valid for weak deformations. This can be seen {in} any consistent gradient expansion scheme, e.g. the semi-classical gradient expansion of the BdG Hamiltonian for superconductors/superfluids, or the Schrieffer-Wolff transformation for Bloch Hamiltonians \cite{WeststromOjanen17, LiangOjanen19}. 

{We conclude that} the Hamiltonian \eqref{eq:HWeyl} has striking similarity to relativistic fermions coupled to non-trivial background geometry or gravity, {albeit with some important caveats}. More precisely, if we consider the low-energy Weyl fermion $\Psi_W$ in terms of the original excitations $\Psi$, {we see}
\begin{align}
\Psi(\vek{x},t) = e^{\im \vek{p}_W \cdot \vek{x}} \Psi_W(\vek{x},t), \label{eq:momentum_rotation}
\end{align}
which, however, corresponds to the anomalous (chiral) rotations in the system, thus making the finite node momentum $p_W$ very important. In the rest of the paper, we will explicitly consider the anomaly implied by \eqref{eq:momentum_rotation} in the presence of non-trivial background fields $e^{\mu}_a(x)$, from Eq. \eqref{eq:Taylor_tetrad}, after reviewing the necessary {background geometry} in the next section. {U(1) gauge fields are assumed to be absent}. We will focus here on $T$-breaking systems, where in the simplest case one finds Weyl nodes of opposite chirality at $\pm \vek{p}_W$, whereas for inversion $P$ breaking systems one has at minimum four Weyl points, which are invariant under $T$ and map non-trivially to themselves under inversion.
 
\subsection{Quasirelativistic fermions}

We briefly summarize quasirelativistic fermions on curved Riemann-Cartan spacetimes here, see e.g. \cite{NiehYan82, ObukhovEtAl97, ChandiaZanelli97, ParrikarEtAl14}. These spacetimes are defined via an orthonormal frame $e^a = e^a_{\mu}dx^\mu$, {giving rise to metric as in \eqref{eq:cone_metric}}, and a (matrix) spin-connection $\hat{\omega}_{\mu} dx^{\mu}$, both of which couple to the Dirac (and Weyl) equations. {Informally}, the $e^a_{\mu}$ is a spacetime ``translation gauge field", while $\hat{\omega}$ is the gauge connection corresponding to local (Lorentz) rotations. See e.g. \cite{NiehYan82b}.

As discussed above and the Introduction, analogous fields arise in the low-energy Weyl Hamiltonian close to the nodes {in condensed matter systems on flat space, giving rise to emergent spacetimes for the low-energy fermions}. These are, however, not strictly relativistic in the sense that the emergent metric does not follow from locally Lorentz invariant spacetimes implied by general relativity, but rather from the microscopic non-relativistic UV theory at low energy. This what we call quasirelativistic and emergent. 
Note that the spin-connection is strictly speaking a gauge field of {a local symmetry entering the Dirac operator. Therefore its emergence needs the corresponding local symmetry}. Notwithstanding, it arises however, e.g. in chiral superconductors and superfluids due to the local combined U(1) symmetry corresponding to gauge and orbital rotation symmetry \cite{LiuCross79, GolanStern18, Nissinen2019}. The tetrad and connection fields give rise to the torsion $T^a=de^a+(\hat{\omega} \wedge e)^a$ and curvature $\hat{R} = d\hat{\omega}_{\mu}-\hat{\omega} \wedge \hat{\omega}$ {field strength} tensors that equivalently characterise the spacetime. From the tetrad one can derive the spacetime metric, which enters as a secondary object, in contrast to usual Riemannian spacetimes where the connection is symmetric and uniquely fixed by the metric.

In terms of equations, the basic quantities are the tetrad $e^a_{\mu}$ and coordinate connection $\Gamma_{\mu\nu}^{\lambda}$. The former is the {metric matrix square-root}
\begin{align}
g_{\mu\nu} = e^a_{\mu} e^b_{\nu} \eta_{ab}, \quad e_{a}^\mu e_{b}^{\nu} \eta_{ab} = g^{\mu\nu} 
\end{align}
by defining a local orthonormal frame, {in terms of $\eta_{ab}= \textrm{diag}(1,-1,-1,-1)$}. Now tensors $X^{a\cdots \mu \cdots}_{b\cdots \nu \cdots}$ can carry local orthonormal (Lorentz) indices and coordinate indices; the two bases can be transformed by contracting with $e^a_{\mu}$ or the inverse $e^{\mu}_a$. The connection consistent with basis changes {defined as $\nabla e^a_{\mu} = 0$}, has two parts, one for local orthonormal indices and one for coordinate indices and is metric compatible. The {connection} determines geometric parallel transport in the system. Without loss of generality it can be written as
\begin{align}
\omega^a_{\mu b} = e^a_{\lambda} e^{\nu}_{b} \Gamma^{\lambda}_{\mu\nu} - e^a_{\nu}\doo_{\mu} e^{\nu}_b \label{eq:spin-connection},
\end{align}
where $\Gamma_{\mu\nu}^{\lambda}$ is the coordinate connection with torsion
\begin{align}
T_{\mu\nu}^{\lambda} = \Gamma^{\lambda}_{\mu\nu} - \Gamma^{\lambda}_{\nu \mu}. 
\end{align}
The connection can be decomposed in terms of torsion as
\begin{align}
\Gamma^{\lambda}_{\mu\nu} = \mathring{\Gamma}^{\lambda}_{\mu\nu} + C^{\lambda}_{\mu \nu},
\end{align}
where $\mathring{\Gamma}^{\lambda}_{\mu\nu} = \frac{1}{2}g^{\lambda\rho}(\doo_{\mu}g_{\nu\rho} +\doo_{\nu}g_{\mu\rho}-\doo_{\rho}g_{\mu\nu})$ is the Christoffel connection fully determined from the metric and $C^\lambda_{\mu\nu} = \frac{1}{2} (T^{\lambda}_{\ \mu\nu} + T_{\mu \ \nu}^{\ \lambda} - T^{\ \ \lambda}_{\mu \nu})$ is the contorsion tensor.

The low-energy quasirelativistic Weyl fermion theory is, in the chiral Dirac fermion basis $\psi = \left(\begin{matrix} \psi_L & \psi_R \end{matrix}\right)^T$, where $\psi_{R,L}$ are Weyl fermions and $\gamma^a = \overline{\sigma}^a \oplus \sigma^{a}$ with $\overline{\sigma}^a = (1,-\sigma^i)$,
\begin{align}
S_{D} = \int d^4 x e~ \frac{1}{2}\overline{\psi}\gamma^a (e^{\mu}_a \im D_{\mu } - p_{Wa})\psi + \hc ~. \label{eq:Dirac_action}
\end{align}
where $e \equiv \det e^a_{\mu}$ and $D_{\mu}$ is the covariant derivative corresponding to the canonical momentum
\begin{align}
D_{\mu} = \doo_{\mu} - \frac{\im}{4} \omega_{\mu}^{ab} \sigma_{ab} - \im q A_{\mu} 
\end{align}
where $\gamma^{ab} = \frac{\im}{2}[\gamma^a,\gamma^b]$ and $A_{\mu}$ is a U(1) gauge potential with charge $q$. They enter the covariant derivative or canonical momentum due to local Lorentz (rotation) and gauge symmetries. For the emergent spin-connection to exist, the local rotation symmetry has to be dynamically generated. See Sec. \ref{sec:chiral} and \cite{Nissinen2019}. Importantly to our applications, the quantity $p_{Wa} = (\mu_W, \vek{p}_W )$ is the shift of the of the Weyl (or Dirac) node at chemical potential $\mu_W = e_0^\nu p_{W\nu}$ and $\vek{p}_{Wa} =  e^i_a p_{Wi}$ in momentum space. The magnitude of latter is a UV-parameter that is fixed (up to small deformations) in the low-energy theory.

\subsection{Anisotropic Newton-Cartan fermions}\label{sec:Newton-Cartan}

A related concept to the {Riemann-Cartan spacetime \eqref{eq:Dirac_action}} is an anisotropic version of a {non-relativistic Newton-Cartan (NC) spacetime. In the latter, we single out a Newtonian time and, in our case, a preferred spatial direction with quadratic dispersion in contrast to the linear Riemann-Cartan case}. In what follows {in} Secs. \ref{sec:chiral} and \ref{sec:WSM}, this preferred direction is {along the Weyl node separation} with uniaxial symmetry and anisotropic scaling. Compared to the standard {NC} case, there is an additional gauge symmetry corresponding to a U(1) number conservation and a local Milne boost symmetry along the anisotropy direction \cite{Son13, BanerjeeMukherjee18, CopettiLandsteiner19, Copetti20}. These will both be gauge fixed to zero and will be applied {mostly} in the case of the chiral superconductor/superfluid, where they are absent naturally for Majorana-Weyl fermions. With the time coordinate fixed, the symmetries of the NC spacetime then correspond to the generalized Galilean transformations $x^i \to x^i +\xi^i(x,t)$ \cite{DuvalEtAl85, Son13, ObersEtAl14, BanerjeeEtAl14, WilsonEtAl20}.

{The} metric is
\begin{align}
g_{\mu\nu} = n_{\mu}n_{\nu} + h_{\mu\nu}
\end{align}
where now $n_\mu$ is a \emph{spacelike} vector, {$e^a_{\mu}$ a (degenerate) tetrad with metric $h_{\mu\nu}$ restricted to the orthogonal subspace}, with $e^0_\mu = \delta^0_\mu$ representing Newtonian time,
\begin{align}
h_{\mu\nu} = \eta^{ab} e^{a}_{\mu}e_{\nu}^b, \quad a,b =0,1,2,
\end{align}
with inverses
\begin{align}
n_{\mu}\ell^{\mu} = 1, \quad e^a_{\mu}\ell^{\mu} =0, \quad e^a_{\mu} e^{\mu}_b = \delta^a_b,\quad  a=0,1,2.
\end{align}
The connection and torsion follow as
\begin{align}
\Gamma^{\lambda}_{\mu\nu} = \mathring{\Gamma}^\lambda_{\mu\nu}[h] + \ell^{\lambda}\doo_{\mu}n_{\nu},
\end{align}
from the condition that $\mathcal{L}_{\ell} h_{\mu\nu} = 0$, equivalent to $\nabla_{\mu}n_\nu = \nabla_{\lambda}h_{\mu\nu}=0$. The torsion is given as
\begin{align}
T^3_{\mu\nu} \equiv n_\lambda T^{\lambda}_{\mu\nu} = -\doo_{\mu}n_{\nu} + \doo_{\nu}n_{\mu} 
\end{align}
and the standard spin-connection perpendicular to $\ell^\mu$, $\mathring{\omega}_{\mu\nu}[h]$, {as in} Eq. \eqref{eq:spin-connection}, amounting to local rotation symmetry along $\ell^\mu$. The fact that $n_{\mu}$ is covariantly constant is natural, since it can be identified with the direction corresponding to non-zero Weyl node separation in e.g. $T$-breaking Weyl systems. 

We discuss in Sec. \ref{sec:chiral} the Landau level problem of Majorana-Weyl fermions corresponding to such a spacetime, with the (right-handed Weyl) action
\begin{align}
S_{W} = \int d^4x \sqrt{g} \psi^\dagger [(\tau^a c_{\perp}e_a^\mu \im \doo_\mu - \tau^3\epsilon(\im \doo_\ell)] \psi + \hc \label{eq:NC_fermion}
\end{align}
where $\epsilon(\doo_\ell) = \doo_\ell^2/(2m)-\mu_F$ in the anisotropic direction with $\doo_\ell = \ell^{\mu}\doo_{\mu}$, corresponding to the non-relativistic dispersion and degenerate metric $\ell^{\mu}\ell^{\nu} = g^{\mu\nu}-h^{\mu\nu}$. In this case the relative anisotropy of the two terms is $c_{\perp}/c_{\parallel} = mc_{\perp}/p_F$, where $p_F = \sqrt{2m\mu_F}$ and $c_{\parallel}=v_F$ the Fermi velocity. This NC model can be matched to the results discussed \cite{Nissinen2019}. Note that a very similar model with Lifshitz anisotropy was considered in \cite{CopettiLandsteiner19}, and the ensuing torsional anomalies for momentum transport in \cite{Copetti20}. For a semimetal under strain, the model {in Sec. \ref{sec:WSM} is correspondingly anisotropic but the precise connection to a specific NC model and symmetries remains to be worked out in full detail}.

\section{Torsional anomalies and Landau levels}\label{sec:torsional_LLs}

\subsection{Torsional Nieh-Yan anomaly}
Now consider Weyl fermions coupled to a tetrad with non-zero torsion and curvature with the U(1) gauge fields set to $A_{\mu} = A_{5 \mu} = 0$, {see however Appendix \ref{sec:appendix_EM}}. As for the U(1) gauge fields, or gravitational fields represented by the metric $g_{\mu\nu}$, the Weyl fermions are anomalous in the presence of non-zero torsion (and curvature).  

We focus on a pair of complex fermions of opposite chirality with currents $j^{\mu}_{\pm}$. The (covariant) torsional anomaly for the axial current $j^{\mu}_5 = j^\mu_{+}-j^\mu_{-}$ is \cite{Yajima96, ObukhovEtAl97, ChandiaZanelli97, Soo99, PeetersWaldron99}
\begin{align}
\doo_{\mu} (e j^{\mu}_5) &= \frac{\Lambda^2}{4\pi^2} (T^a \wedge T_a - e^a \wedge e^b \wedge R_{ab}) \label{eq:NYanomaly}\\ 
& + \frac{1}{192\pi^2}\tr(R\wedge R) \nonumber\\
= \frac{\Lambda^2}{4\pi^2} &\epsilon^{\mu\nu\lambda\rho}(\frac{1}{4}T^a_{\mu\nu}T_{a\lambda\rho} - \frac{1}{2}e^a_{\mu}e^b_{\nu}R_{ab\lambda\rho}) + O(\doo^4). \nonumber
\end{align}
For a discussion of the relativistic {torsional} anomaly term, we refer to \cite{NiehYan82, NiehYan82b, ObukhovEtAl97, ChandiaZanelli97, Comments}, and for applications in topological condensed matter systems, \cite{SunWan14, ParrikarEtAl14, FerreirosEtAl19, Nissinen2019, Stone2019, Copetti20, LiangOjanen19b}. For the mixed terms between torsion and U(1) gauge potentials, see e.g. \cite{KubotaEtAl01}. {We focus on the anomaly contribution solely due to the geometry (tetrads), we will not consider them}. Ref. \cite{FerreirosEtAl19} also considered novel ``axial" tetrads $e^a_{\mu R} \neq e^a_{\mu L}$ at two Weyl nodes $R,L$, with (vector like) $T^5$ appearing as in Eq. \eqref{eq:U1_anomaly_eqs}. We will require $e_R = \pm e_L$ but this is actually rather strong constraint basically only allowing for (improper) rotations that can be gauged away. In the chiral Weyl superfuid/conductor or minimal time-breaking semimetal, $e_R =-e_L$ but this just the chirality of the nodes and is built in the axial nature of torsion. Intriguingly the trace part of torsion arises as the gauge field of local Weyl scalings but this comes, since non-unitary, with a complex gauge coupling \cite{NiehYan82}. {The presence of different (chiral) tetrad couplings and overall symmetry considerations would be highly interesting for e.g. parity breaking and other non-minimal Weyl systems with several nodes, some of which coincide in momentum space}.

{To conclude}, we note the following salient properties related to the NY anomaly term: i) Despite appearances, it is given by the difference of topological terms, albeit in five dimensions \cite{ChandiaZanelli97}. ii) The NY anomaly term is of second order in gradients and therefore the leading contribution from the background geometry in linear response. iii) The UV-cutoff is isotropic in momentum space by (local) Lorentz invariance but is multiplied by the geometric term, which can be anisotropic. {In condensed matter applications, we do not expect Lorentz invariance so in principle non-isotropic anomaly coefficients can arise (see e.g. Sec. \ref{sec:thermal})}. iv) The NY term has contributions from the torsion and curvature, dictated by local exactness $d(e^a \wedge T_a) =  T^a \wedge T_a -e^a\wedge e^b \wedge R_{ab}$. The two contributions are a priori independent before the geometry (the torsionful connection) is fixed. The anomaly is therefore physical input for the spacetime geometry or connection \cite{Nissinen2019}. In more pragmatic terms, the anomaly coefficient $\Lambda^2$ can be computed in the case when $\hat{\omega}_\mu = 0$, although the constraints of a consistent spacetime geometry should be kept in mind.

\subsection{Quasirelativistic fermions and torsional Landau levels}
Now we proceed to compute the torsional NY anomaly in non-relativistic systems utilizing the Landau level argument. To set the stage and remove confusions before presenting our main results, we briefly review (quasi)relativistic torsional Landau levels {with linear spectrum}, see e.g. \cite{ParrikarEtAl14}. The computation of the Landau levels is close to and inspired by the spectral flow obtained in \cite{Volovik85, BalatskiiEtAl86} for momentum space gauge fields at $p_W\neq 0$. Similar considerations for $p=0$ can be found in \cite{Stone2019, Stone2019b}.

The Weyl particles are governed by the effective Hamiltonian
\begin{align}
H_{\rm W} = \sigma^a e^{i}_{a}(\im \doo_i - p_{W,i}) + \textrm{h.c.}
\end{align}
where $\vek{p}_W$ is the location of the Weyl point. Due to the lack of protecting symmetries (namely at least broken $P$ or $T$) the shift vector
\begin{align}
p_{W,\mu} = (\mu_W, \vek{p}_W)
\end{align}
is necessarily non-zero for the existence of the Weyl point. However, we will focus on the $T$-breaking case with two nodes of opposite chirality at $\pm\vek{p}_W$ and assume that $\mu_W$ is zero unless otherwise specified.

 In this section, we assume that the coordinate dependence of the Hamiltonian arises solely from the tetrad $e^\mu_a(x)$, while the location of the node, {$p_{aW}$ is assumed to be constant}. Note that the coordinate momentum $p_{W\mu} \equiv e^a_\mu p_{Wa}$ can still vary and in the case $T^a_{\mu\nu} \neq 0$ there is non-zero torsion. Torsional LLs arise when, say, $\frac{1}{2}\epsilon^{ijk}T^3_{jk} = T_B\unitvec{z}^i$ is constant with the other torsion components and spin connection vanishing. We discuss later in Secs. \ref{sec:chiral}, \ref{sec:WSM} on how to make the identification between low-energy emergent gravitational fields and microscopic background fields in specific examples. 
 
\subsubsection{Torsional Landau levels}

Specifically, the {assumed} (semi-classical) tetrads $e^a = e^a_{\mu}  dx^{\mu}$ and the inverse $e_a = e^{\mu}_a \doo_{\mu}$ are, following \cite{Volovik85, BalatskiiEtAl86, ParrikarEtAl14},
\begin{align}
e^0 &= dt, \quad e^{1} = dx, \quad e^{2} = dy, \quad e^3 = dz-T(y)dx \nonumber \\
e_0 &= \doo_t,\quad e_{1} = \doo_x+T(y)\doo_z, \quad e_2 = \doo_y, \quad e_3 = \doo_z .\label{eq:torsion_tetrad}
\end{align}
Now we compute the spectrum of the Weyl fermions in the presence of a constant torsional magnetic field $T(y)=T^3_B y$. The corresponding metric is
\begin{align}
g_{\mu\nu}dx^{\mu}dx^{\nu} &= \eta_{ab}e^a e^{b} \nonumber \\
&= dt^2-(1+T(y)^2)dx^2-dy^2\\
&\phantom{=}-2 T(y) dx dz-dz^2 . \nonumber
\end{align}
The torsion is given as $T^3_{ij} = \doo_\mu e^3_\nu-\doo_{\nu}e^3_{\mu}$ or $T^3 = de^3 =\frac{1}{2} \doo_y T(x) dx \wedge dy$, i.e. $T^3_{xy} = -\doo_y T(y) =T_B^3$. In analogy with the electromagnetic tensor, we will call $\frac{1}{2} \varepsilon^{ijk} T_{jk}^a$ and $T^a_{0i}$ torsional magnetic and electric fields, respectively.

\begin{figure}
\centering
\includegraphics[width=220pt]{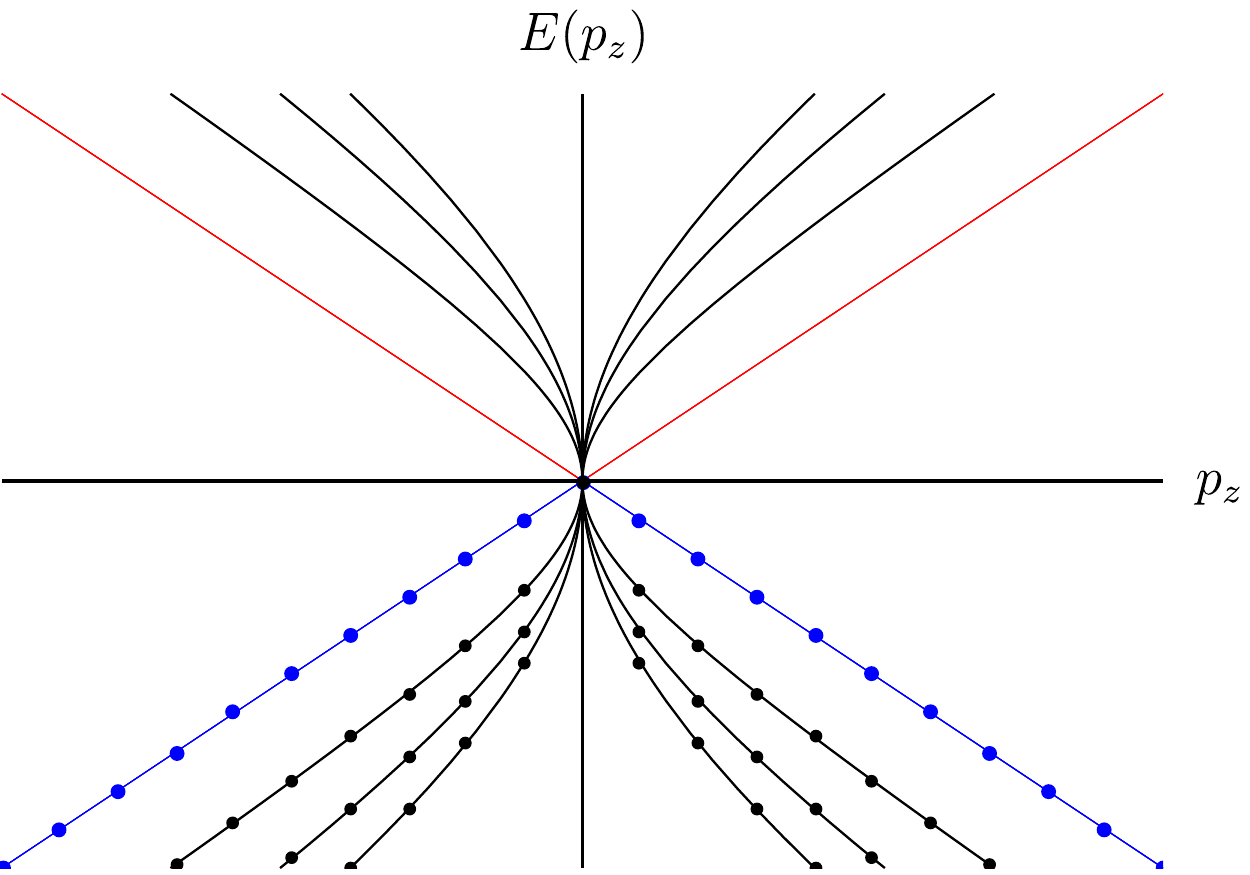}
\caption{Dispersion of left-handed (LLL in blue) and right-handed Weyl fermions (LLL in red) at $p_W=0$ under a torsional magnetic field, respectively.}
\label{fig:relativistic_TLL}
\end{figure}

The Weyl hamiltonian couples to the non-trivial vierbein as, $\chi$ being the chirality,
\begin{align}
\label{eq:hamT}
    H_\ch =& \frac{\ch}{2} \sigma^a e_a^i  \khat_i +\hc \nonumber \\
    =& \ch\begin{bmatrix}\mom_z && \khat_x+\khat_z \tb y - i\khat_y\\ \khat_x+\khat_z\tb y + i\khat_y && -\hat{p}_z \end{bmatrix}. 
\end{align}
As usual, the energy eigenvalues are obtained from squaring the Hamiltonian 
\begin{align*}
                H^2 &= \sigma^a e^i_a \khat_i e^j_b
                \sigma^b \khat_j  = e^i_a e^j_b \sigma^a \sigma^b \khat_i \khat_j + e^i_a\sigma^a\sigma^b \{\khat_i,e^j_b\}\khat_j \\
                & =  e^i_a e^j_b (-\eta^{ab}+i\epsilon^{abc}\sigma^c) \khat_i \khat_j + \frac{i\tb}{2}[\sigma^2,\sigma^1] \khat_z \\
               &  = -g^{ij}\khat_j\khat_j - \tb\sigma_3\khat_z.
               \\
                & = \khat_y^2 + \khat_z^2 + (\khat_x + \tb \hat{y}\khat_z)^2 - \tb\sigma_3 \hat{p}_z.
                \end{align*} 
We see \eqref{eq:hamT} is equivalent to a LL problem in a magnetic field [Eq. \eqref{eq:Hmag} for \(B^z = \tb\) and \(e = p_z\) in Appendix \ref{sec:appendix_EM}]. With those identifications, the spectrum is consequently [from Eq. \eqref{eq:relEMspectrum}]:
\begin{align}
\label{eq:tllspectrum}
    E(p_z) = \begin{cases}\pm \sqrt{p_z^2+2|p_z\tb |n}, \quad n\geq1 \\ \text{sgn}(\tb )\ch|p_z|, \quad n = 0. \end{cases}
\end{align} 
The lowest Landau level (LLL) is chiral and unpaired with the simple eigenfunctions, $\sigma^3=\pm1$,
\begin{align}
\Psi_{\sigma^3}(x,p_x,p_z) \sim e^{\im (p_x x+p_z z)} e^{\pm(p_x y-p_z T_B y^2/2)} \label{eq:LLL_gaussian}
\end{align}
where the (pseudo)spin or helicity is determined by $\sgn(p_zT_B)$. We stress that the shape of the spectrum is in general also modified due to the momentum replacing the electric charge:  left-handed states now disperse as \(E<0\) and right-handed states as \(E>0\) (or vice versa, depending on the sign of the field), see Fig. \ref{fig:relativistic_TLL}.

\subsubsection{Spectral flow and anomaly}

\begin{figure}
    \centering
\includegraphics[width=.49\textwidth]{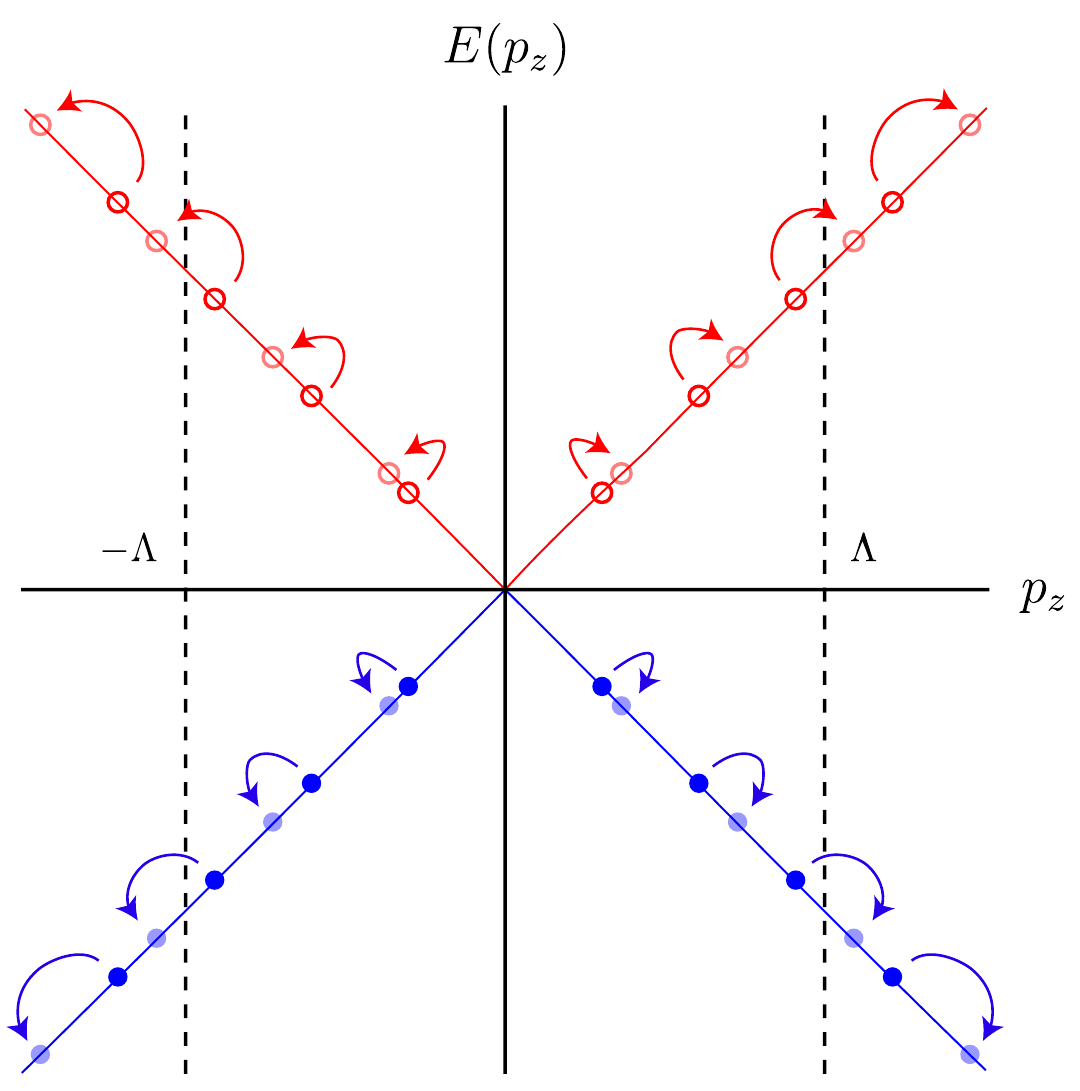}
    \caption{Relativistic spectral flow at $k=0$ in the presence of torsion, with the adibatic transfer of states. Dashed line indicates the location of the cutoff $\Lambda$.  }
    \label{fig:relativistic_spectral_flow}
\end{figure}

Analogously to the Landau level calculation with electromagnetic fields, we may turn on a constant torsional electric field parallel to \(\tb \) by introducing time-dependence to the vierbein as \(e_z^3 = 1+\te t\) where $\te t \ll 1$. Then we have $e^z_3 = (1+\te t)^{-1} \approx 1-\te t$. This induces adiabatic time-dependence $\doo_t p_z = (\doo_t e^3_z) p_3$, analogous to the Lorentz force, which leads to spectral flow of states through the momentum dependent torsional electric field. The number currents, in the vicinity of the node $p_z =  e^3_z p_3 = p_{Wz}=0$ are for both chiralities
\begin{align}
\label{eq:tllcurrent}
   e j^0_\chi(t) &= \frac{\tb }{2\pi} \int_{-\Lambda}^{\Lambda}\frac{dp^3}{2\pi}|p_z| \nonumber \\
    &= - \Lambda^2\frac{\tb (1+\te t)}{4\pi^2} = -\Lambda^2\frac{T^3_{xy} e_z^3}{4\pi^2},
\end{align}
where a cutoff \(\Lambda\) has been introduced to regularize the momentum dependent current density {and spectrum}. We see that for $E<0$, particles flow below the cutoff, whereas for $E>0$, holes flow above the cutoff, see Fig. \ref{fig:relativistic_spectral_flow}. Then, taking into account the fact that the tensorial current density is modified by the volume element $e d^4 x$ in the presence of torsion, see e.g. \cite{Soo99, BradlynRead15},
\begin{align}
    \dot{e j^0_{\chi}} &= \mp \Lambda^2\frac{T^3_{xy} \part e_z^3}{4\pi^2} = \mp \Lambda^2\frac{\tb \te }{4\pi^2} \nonumber\\
    &= \mp\frac{\Lambda^2}{32\pi^2}\levic T^3_{\mu\nu} T^3_{\rho\sigma}, \label{eq:spectral_flow_anomaly}
\end{align}
from holes or particles moving above or below the cutoff, respectively, depending on the direction of the torsional electric field. This is the vacuum regularization that was {also} used in Ref. \onlinecite{ParrikarEtAl14} in the sense $n_{\rm vac} =\sum_{\abs{E_n}\leq \Lambda} \sgn(E_n)$, where an additional factor of one half was present, presumably due to comparison with anomaly inflow from five dimensions. Generalizing this to a fully covariant expression, see the Appendix \ref{sec:appendix_EM}, gives
\begin{align}
    \frac{1}{e}\partial_\mu(ej^\mu_{5}) =  \frac{1}{e}\frac{\Lambda^2}{16\pi^2}\levic T^3_{\mu\nu} T^3_{\rho\sigma},  \label{eq:j_spectral_flow}
\end{align}
and in particular $\doo_{\mu} (ej^\mu)=0$ as required. We discuss {the relativistic vacuum and the spectral flow leading to \eqref{eq:j_spectral_flow}}, as compared to nodes at finite momenta and axial U(1) fields, more in the next section.

\subsubsection*{Torsional anomaly for \(p_W \neq 0\)}

If we now displace the Weyl nodes in the relativistic case \eqref{eq:hamT} by \(p_z = \pm p_{W}\) in momentum space, corresponding to a $T$-breaking Weyl system, the spectrum \eqref{eq:tllspectrum} takes the form
\begin{align}
    E(p_z) = \begin{cases}\pm \sqrt{(p_z\pm 
    p_{W})^2+2|p_z\tb |n}, \quad n\geq1 \\ \text{sgn}(\ch p_z\tb )(p_z\pm p_{W}), \quad n = 0. \end{cases}
\end{align}

\begin{figure}[h]
    \centering
     \includegraphics[width=250pt]{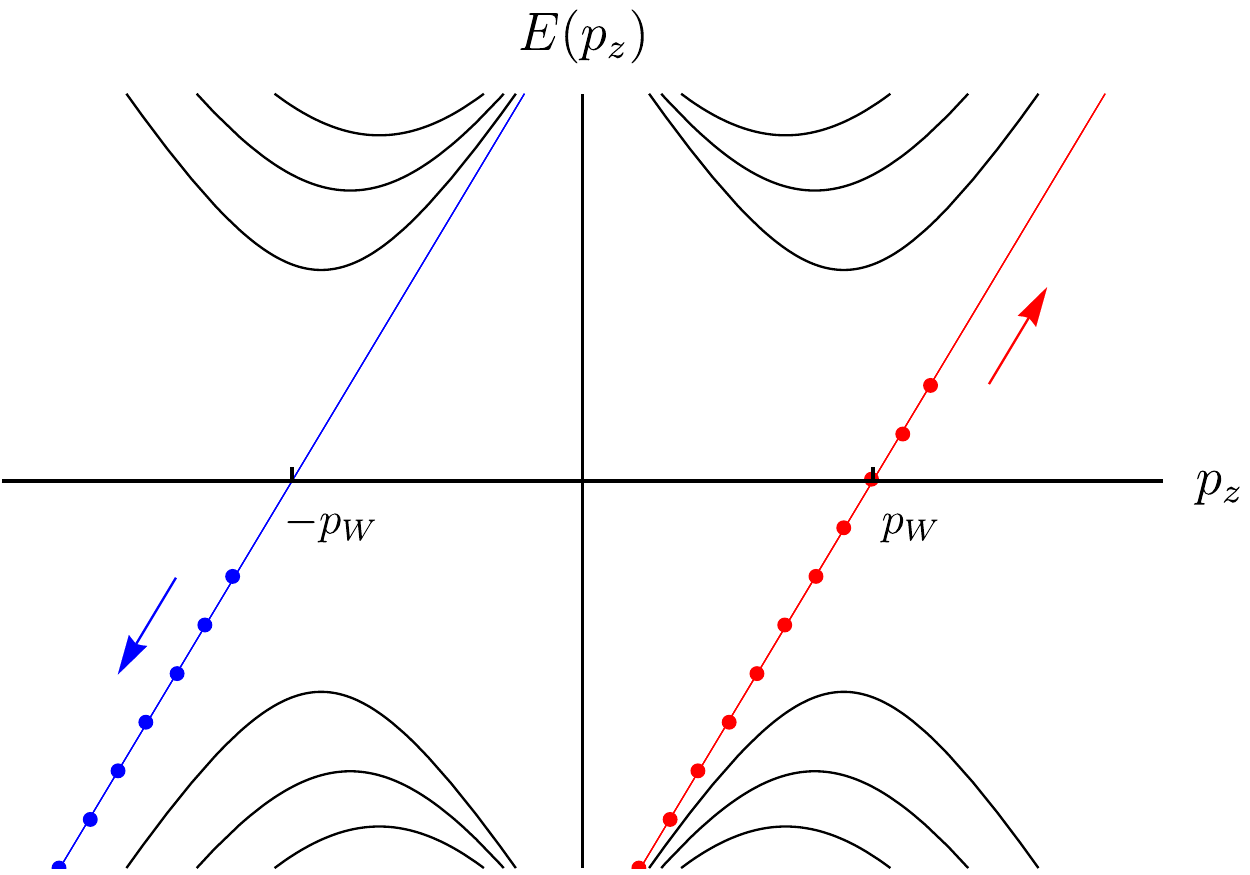}
    \caption{Left-handed Weyl particles at $k_z = k_0$ (LLL in red) and right-handed Weyl holes at $k_z = -k_0$ (LLL in blue) under a torsional magnetic field. Spectral flow is indicated with the arrows.}
    \label{fig:pseudotorsion}
\end{figure}

The lowest, chiral Landau level looks exactly like that of a Weyl fermion in an axial magnetic field, Eq. \eqref{eq:displacedham}. Higher levels are distorted due to the effective charge carried by the particles being their momentum. See Fig. \ref{fig:pseudotorsion}.

Since the node is at finite momentum $p_W\neq 0$, also the spectral flow summation is centered around $p_W \pm \Lambda'$, {where $\Lambda'$ is a cutoff from e.g. the validity of the linear spectrum}. For notational convenience and comparison to Eq. \eqref{eq:j_spectral_flow}, we introduce the momentum cutoff as $\Lambda' = \frac{\Lambda_{\rm rel}^2}{2} p_W$, where we expect $\frac{\Lambda_{\rm rel}^2}{2} \ll 1$, this being the dimensionless ratio of the cutoff of the linear spectrum to $p_W$. The spectral flow results in the expression, where particles and holes simply add at the two nodes,
\begin{align}
    \frac{1}{e}\partial_\mu(ej^\mu_{5}) = \frac{1}{e}\frac{p_W^2 \Lambda_{\rm rel}^2}{16\pi^2}\levic T^3_{\mu\nu} T^3_{\rho\sigma} 
\end{align}
which shows that the NY anomaly cutoff is proportional to the node momentum $p_W$, and is small by a factor $\Lambda^2_{\rm rel}\ll 1$ corresponding to the validity of the linear Weyl approximation.

\subsubsection{Comparison of torsion to U(1) fields}

From Figs. \ref{fig:relativistic_TLL} and \ref{fig:pseudotorsion}, we see that the spectrum of torsional LLs resemble the LL spectrum of charged particles in U(1) axial and vector fields, with the momentum dependent charge to torsion kept in mind. {See appendix \ref{sec:appendix_EM} for a complete review of the U(1) case for comparison.} It is well-known that the contribution of torsion for complex chiral Weyl fermions can be equivalently cast in terms of the axial gauge field $\gamma^5 S^{\mu} \equiv \gamma^5 \varepsilon^{\mu\nu\lambda\rho} T_{\nu\lambda\rho}$ corresponding to the totally antisymmetric torsion, see e.g. \cite{ChandiaZanelli97, Soo99}. We stress that while the spectral equivalence of torsional and U(1) LLs is of course expected, the physical appearance of the anomaly is drastically different: the density of states of the LLs depend on momentum and thus the dimensional coefficient $\Lambda^2$ and the need for an explicit UV-cutoff appears. {Similarly, the physics of Figs. \ref{fig:relativistic_spectral_flow} and \ref{fig:pseudotorsion} is completely different, although both arise from spectral flow in momentum space under torsion.}

On this note, although the relativistic result in \eqref{eq:spectral_flow_anomaly} is familiar, there seems to be still confusion in the literature about the role of torsional Landau levels in momentum space and the validity of the NY anomaly due to the explicit UV cutoff. For relativistic Weyl fermions with Lorentz invariance up to arbitrary scales, the spectral flow is symmetric around $p=0$, leading to the conclusion that the anomaly indeed can cancel. This is simply by the observation that, in the absence of Lorentz symmetry breaking at high energy, no net transfer of occupied and empty states in the vacuum takes place during the adiabatic spectral flow, cf. Fig. \ref{fig:relativistic_spectral_flow}. The net transfer of $j_5$ requires left-right asymmetric regularization at the scale of $\Lambda$ with chirality disappearing above that scale, maintaining $\doo_{\mu}j^{\mu}=0$ \cite{ParrikarEtAl14}. Alternatively, at the very least, there is a divergence as $\Lambda\to\infty$. In contrast, for quasirelativistic Weyl fermions at finite node momentum and an explicit cutoff to the Weyl spectrum, the spectral flow can terminate due to the non-relativistic corrections at the cutoff scale of $\Lambda^2_{\rm rel}$, {also implying that chirality is no longer well-defined}, leading to net transport of states and momenta relative to the vacuum (and other quantum numbers of the Weyl fermions if present). {A related fact is that the momentum that plays the role of chirality, which remains physically well-defined irrespective of the scale}. {We also note that the flow is composed of particles and antiparticles (holes) at the different nodes}. It would be interesting to study the {detailed} role of the breakdown of relativistic spectrum and spectral flow numerically, following Ref. \onlinecite{SukhachovEtAl18}. {There only the charge density at finite chemical potential from the node is analyzed, corresponding to Fig. \ref{fig:B5E} and the expected deterioration away from the Weyl node is verified.}

\section{Chiral Weyl superfluids and superconductors}\label{sec:chiral}
Now we discuss the role of the torsional anomaly in $p$-wave superfluids and superconductors with gap nodes and associated Weyl-Majorana quasiparticles \cite{Volovik84, Volovik90, VollhardtWoelfle, ReadGreen00, PalumboPachos16, MaranerPachosPalumbo18}. Close to the nodes, the Fermi energy is tuned to the Weyl point due to the existence of the $p+ip$ pairing amplitude. The chiral anomaly is related to the non-conservation of momentum in the condensate and normal state quasiparticles \cite{BevanEtAl97}. The relation of {this to the} torsional gravitational anomaly and the LL spectral flow was briefly pointed out in Ref. \cite{Nissinen2019}. Earlier related work can be found in \cite{Volovik85, Volovik1986b, BalatskiiEtAl86, CombescotDombre86, Volovik90, KobayashiEtAl18, IshiharaEtAl19}. 

The spinless gap amplitude, with equal spin pairing understood, takes the form
\begin{align}
\Delta(\vek{p}) = \frac{\Delta_0}{p_F} (\unitvec{m}+\im \unitvec{n}),
\end{align}
where $c_{\perp} = \Delta_0/p_F$ has units of velocity. The direction $\unitvec{l}= \unitvec{m}\times \unitvec{n}$ is a low-energy Goldstone variable for the condensate. At low-energy, the direction of $\unitvec{l}$ can fluctuate and there is combined U(1) gauge symmetry \cite{LiuCross79} in the $\unitvec{m}-\unitvec{n}$ plane, leading to the Mermin-Ho relations between $\unitvec{l}$ and $\vek{v}_s$ \cite{MerminHo76, VollhardtWoelfle, Volovik03}. In the following, we focus on the Landau levels and torsion, keeping the magnitudes of $p_F$ and $\Delta_0$ fixed. {Related to this, for the superconductors, the end results apply the case where the EM potential $A_{\mu}=0$} which amounts to the case where we work in the gauge where $\mathbf{v}_s - \vek{A} \to \mathbf{v}_s$. In the following {computations} we will set $\mathbf{v}_s = 0$ as well, since this corresponds to the case where one has only torsion, see Ref. \onlinecite{Nissinen2019} for the general case {with superfluid velocity}. The orientation of the orthonormal triad $\unitvec{l}$ can still rotate {for the torsional textures}.

Considering {first} the simple homogenous case, the linearization of the BdG Hamiltonian takes the form of a Weyl Hamiltonian close to the nodes of $E(\vek{p})$ at $\vek{p}=\mp p_F\unitvec{l}$,
\begin{align}
H_{\rm BdG}(\hat{\vek{p}}) &= \left(\begin{matrix}  \epsilon(\hat{\vek{p}}) & \frac{1}{2}\{\hat{\vek{p}},\Delta(\vek{p})\} \\ \frac{1}{2}\{\hat{\vek{p}},\Delta^{\dagger}(\hat{\vek{p}})\} & -\epsilon(-\vek{p})\end{matrix}\right) \\
&\approx \pm \tau^a e^i_a(p_i \mp p_{F,i}) .\nonumber
\end{align}
Note that the BdG excitations are Majorana, $\Phi^{\dagger}(\vek{p}) = \tau^1 \Phi(-\vek{p})$, as expected in a BCS paired system. Here we have taken the normal state dispersion $\epsilon(\vek{p}) = \frac{p^2-p^2_F}{2m}$, where $m$ is the $^3$He atom mass. The tetrads are
\begin{align}
e^i_1 = c_{\perp}\unitvec{m}, \quad e^i_{2} = -c_{\perp} \unitvec{n},\quad  e^i_{3} =- c_{\parallel}\unitvec{l}, \label{eq:3HeA_tetrad}
\end{align}
where $c_{\parallel} \equiv \frac{p_F}{m} = v_F$. Henceforth, to conform with relativistic notation, we will work with dimensionless tetrads in units of $c_{\parallel} = 1$. The quasiparticle dispersion is $E(\vek{p})=\pm \sqrt{\epsilon(\vek{p})^2 + \vert\Delta(\vek{p})\vert^2} \approx \pm \sqrt{c_\parallel q_{\parallel}^2+c_{\perp}^2 q_{\perp}^2}$, with $\vek{q} = \vek{p}-\vek{p}_F$ for the Weyl quasiparticles. The linear expansion is valid when $\abs{\vek{p}-\vek{p}_F} \ll p_F$ which provides an explicit cut-off for the Weyl description, requiring that the remainder
\begin{align}
\frac{1}{2}\frac{\doo \epsilon(\vek{k})}{\doo k^i \doo k^j} (p-p_F)^i (p-p_F)^j = \frac{1}{2m} (\vek{p}-\vek{p}_F)^2 \\
\ll e_a^i (\vek{p}-\vek{p}_F)_i .\nonumber
\end{align}
This leads to the condition, in addition to the trivial $\vert \vek{p}-\vek{p}_F\vert \ll p_F$ from the Taylor expansion of $\epsilon(\vek{p})$, that
\begin{align}
E_{\rm Weyl} \ll m c_{\perp}^2 = \left(\frac{c_{\perp}}{c_{\parallel}}\right)^2 E_{F}. 
\end{align}
which will prove important later. In particular, the energy cutoff for the Weyl quasiparticles is anisotropic in momenta $\vek{q} = \vek{p}-\vek{p}_F$ around the Weyl point,
\begin{align}
q_{\perp} \ll \left( \frac{c_{\perp}}{c_{\parallel}} \right)p_F, \quad q_{\parallel} \ll \left( \frac{c_{\perp}}{c_{\parallel}} \right)^2p_F, \label{eq:Weyl_momenta}
\end{align}
if we consider the Weyl fermion system in the case where the background fields couple parallel and perpendicular directions \cite{Nissinen2019}. {This happens in the chiral system since the three direction are coupled by $\unitvec{l} = \unitvec{m} \times \unitvec{n}$ and the corresponding Mermin-Ho relations.}

\begin{figure}
\centering
\includegraphics[width=200pt]{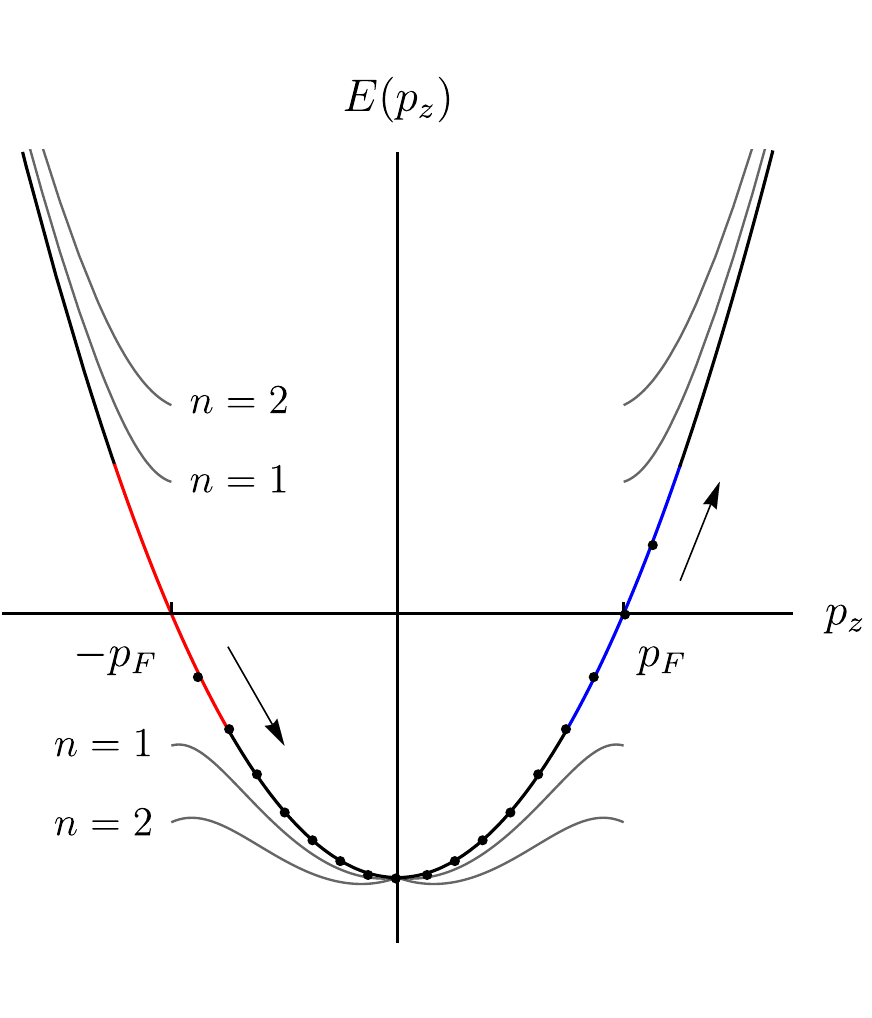}
\caption{The torsional LL spectrum for the anisotropic Newton-Cartan model in chiral superfluids/conductors with the spectral flow indicated. Note that we have inverted the hole-like right-handed Landau level at $-p_F$ and the spectrum is particle-hole doubled. Overall there is a corresponding factor of 2 from spin-degeneracy.}
\label{fig:quadratic_spectrum}
\end{figure}

\subsection{Landau levels in linear approximation}
To compute the LL levels in the order parameter texture corresponding to a torsional magnetic field, we can take the "weak-twist" texture \(\m + i\n = \x + i\y - iT_Bx\z\) with \(|Bx| \ll 1\), which corresponds to $\l = \z + T_Bx\y$\cite{Volovik85, BalatskiiEtAl86, CombescotDombre86}. The BdG Hamiltonian then takes the form
\begin{align}
   H_{\rm BdG}& =  \begin{bmatrix}
    \epsilon(\hat{\vek{p}}) & \half \{\gap ^i,\khat_i\}\\
     \half \{\gapc\phantom{.}^i,\khat_i\}& -\epsilon(-\hat{\vek{p}}) 
    \end{bmatrix} 
  \\ =&  \begin{bmatrix}
    \epsilon(\hat{p}_x, p_y, p_z)  &  \frac{\gap_0}{\pf}[\mom_x + i(p_y-T_Bp_z x)]\\
      \frac{\gap_0}{\pf}[\mom_x - i(p_y-T_Bp_z x )]& -\epsilon(-\hat{p}_x,-p_y, -p_z) 
    \end{bmatrix}. \nonumber
\end{align}
Near the gap node $\vek{p} = -\pf\l$ we may linearize the operator $\epsilon(\hat{\vek{p}})$ as $\eps \approx -v_F\l \cdot(\hat{\vek{p}} + \pf\l) \approx -v_F(p_z+p_F)$. This leads to 
\begin{align}
H_+ = e^i_a\tau^a(p_i-p_F e_i^3) = \tau^a (e^i_a \hat{p}_i - p_F\delta^3_a)
\end{align}
with
\begin{align}
    e^i_a = (c_\perp\delta^i_1, -c_\perp[\delta^i_2-T_Bx\delta^i_3], -c_\parallel\delta^i_3),
\end{align}
where we remind that \(c_\parallel \equiv v_F\) and \(c_\perp \equiv \frac{\Delta_0}{\pf}\). This corresponds, up to the sign of the field $T_{B}$ and the tetrad, to the case \eqref{eq:torsion_tetrad} after a rotation in the $x-y$ plane. 

After moving to scaled coordinates $c_\perp^{-1} x \equiv \Tilde{x}$, $c_\perp^{-1} y \equiv \Tilde{y}$, $c_\parallel^{-1}z \equiv \Tilde{z}$, corresponding to dimensionless and scaled momenta \(p_a \equiv e^i_ap_i\), we can define the annihilation operator \(\hat{a} \equiv \frac{1}{\sqrt{2|T_Bp_z|}}\left[(| T_Bp_z|\Tilde{x} - p_{\Tilde{y}}) + i\mom_{\Tilde{x}} \right]\) to arrive at the Hamiltonian
\begin{align}
   H_{p_z<0} =  \begin{bmatrix}
    p_3+\pf & \sqrt{2|T_Bp_z|}i\hat{a}^\dagger\\
      -\sqrt{2|T_Bp_z|}i\hat{a} & -(p_3+\pf) 
    \end{bmatrix}, \label{eq:H_negative}
\end{align}
which is \eqref{eq:Hmag} after a Galilean boost \(p_3 \to p_3 + \pf\). The eigenstates are then
\begin{equation}
    \Psi_{n,p_z<0}  = \begin{pmatrix}u_n \phi_n \\ v_n \phi_{n-1}\end{pmatrix}e^{i(p_zz+p_yy)}.
\end{equation}
where $\phi_n \equiv \phi_n(x)$, for $n\geq0$, are harmonic oscillator eigenstates and vanish otherwise. The condition for normalization is \(|u_n|^2 + |v_n|^2 = 1\), corresponding to the BdG particle and hole amplitudes.
Carrying out a corresponding calculation at the Weyl point $\vek{p} = \pf\l$, we have the Hamiltonian
\begin{equation}
   H_{p_z>0} =  \begin{bmatrix}
    p_3-\pf & -\sqrt{2|T_Bp_z|}i\hat{a}\\
      \sqrt{2|T_Bp_z|}i\hat{a}^\dagger & -(p_3-\pf) 
    \end{bmatrix}, \label{eq:H_positive}
\end{equation}
which can be identified as the left-handed Hamiltonian \(H_- = -e^i_a\tau^a p_i\) after a rotation about \(\l\) such that \(\m \to -\m\) and \(\n \to -\n\).
Its eigenstates are
\begin{equation}
    \Psi_{n,p_z>0} = \begin{pmatrix}u_n \phi_{n-1} \\ v_n \phi_{n}\end{pmatrix}e^{i(p_zz+p_yy)}.
\end{equation}
Depending on the chirality, i.e. sign of momentum at the node, the LLL is either particle- or holelike {as in Eq. \eqref{eq:LLL_gaussian}}. The conclusion is that the spectrum looks like the relativistic spectrum in Fig. \ref{fig:pseudotorsion}, when the linear approximation for $\epsilon(\vek{p}) \approx \pm c_{\perp}(p_z-p_F)$ is valid, Eq. \eqref{eq:Weyl_momenta}. This corresponds to the spectrum of axial U(1) fields with momentum dependent charge and density of states per LL. The density of states is \eqref{eq:dos} in the scaled coordinates, which gives, with $e^0_{\mu} = \delta^0_{\mu}$,
\begin{align}
    j^0 dV = e j^0 d\Tilde{V}= \frac{|p_zT_B|}{4\pi^2} d\Tilde{V}.
\end{align}

\subsection{Anisotropic Newton-Cartan model}
We just showed that the simple order parameter texture in chiral superfluid or superconductor gives rise to the torsional LLs for the low-energy Weyl quasiparticles, in the linear regime close to nodes. We can however consider quadratic dispersion beyond the linear approximation
\begin{align}
\epsilon(\vek{p}) = \frac{\vek{p}^2}{2m}-\mu_F \to \frac{p_z^2}{2m} -\mu_F, \label{eq:NC_dispersion}
\end{align}
which corresponds to the anisotropic Newton-Cartan (Majorana-Weyl) fermion model in Sec. \ref{sec:Newton-Cartan}. 

The above model has the same regime of validity in the chiral superfluid or superconductor as the linear approximation in Eq. \eqref{eq:Weyl_momenta}, since it also neglects the rotationally invariant dispersion $\epsilon(\vek{p})$ of the normal state, {see also Ref. \onlinecite{Nissinen2019}}. The chiral $p$-wave BCS state has the uniaxial anisotropy of Eq. \ref{eq:NC_dispersion}, however, and this carries to the low-energy Weyl description in the form of the emergent spacetime. The other benefit of the anisotropic model \eqref{eq:NC_dispersion} is that the LL spectrum can be computed for momenta far from $p_F$, up till $p=0$, corresponding to the filled levels of the non-relativistic Fermi system, which are absent in the relativistic linear model. {This is important for the global properties of the chiral spectrum and anomaly}. In this way the contribution to the anomalous current from the superfluid vacuum can be analyzed, see Sec. \ref{sec:vacuum_current}.

The spectrum follows simply from Eqs. \eqref{eq:H_negative}, \eqref{eq:H_positive} by the substitution $\mp(p_3\pm p_F) \to \pm\epsilon(\pm p_z)$. From squaring the Hamiltonian, the corresponding eigenvalues are at both nodes
\begin{align}
E_n &= \pm\sqrt{\epsilon(p_z)^2+c_\perp^2|T_Bp_z|2n}, \nonumber \\
E_0 &= \pm \sgn(p_zT_B) \epsilon(p_z).
\end{align}
for \(n\geq 1\). The LLL state retains the gaussian form \eqref{eq:LLL_gaussian}. The condition for normalization is \(|u_n|^2 + |v_n|^2 = 1\), and consequently the particle and hole amplitudes are in both cases
\begin{equation}
    u_n = \sqrt{\frac{E_n+\epsilon(p_z)}{2E_n}}, \qquad v_n = i\sqrt{\frac{E_n-\epsilon(p_z)}{2E_n}}.
\end{equation}
With $E_0 = \epsilon(p_z)$ we have $v_0 = 0$, meaning that the lowest level particles appear only for \(p_z < 0\). For \(p_z > 0\) \(u_0 = 0\) when \(E_0 = -\epsilon(p_z)\), so for positive momenta only holes appear at the lowest level, as we found for the linear model. In this case we must, however, remember that the hole spectrum arises due to the Majorana doubling of the BdG spectrum and is not physical. This cancels with a corresponding factor of two from spin-degeneracy in the Fermi system. This leads to the LL spectrum in Fig. \ref{fig:quadratic_spectrum}.

\subsection{Spectral flow, axial density and consistent anomalous vacuum current} \label{sec:vacuum_current}

Now we are equipped to compute the spectral flow resulting from torsional Landau levels, corresponding to the covariant torsional NY anomaly. For the anisotropic Newton-Cartan model we can also compute the consistent vacuum current of the condensate, since the dispersion takes into account the filled states below the Fermi-level which is not the case for the linear approximation close to the Weyl nodes. {For the chiral superfluid (or -conductor) we have to take into account that the particles are Majorana-Weyl but a factor of two results from the spin-degeneracy}. 

\subsubsection{Axial density}

The {torsional  spectral flow leads to the anomalous density} as
\begin{align}
e j^{0}_{\pm} = \int_{\mp p_F - \frac{p_F \Lambda^2}{2}}^{\mp p_F + \tfrac{p_F \Lambda^2}{2}} dp^3 N_{\rm LL}(p_z) = \pm \frac{p_F^2(\frac{c_{\perp}}{c_{\parallel}})^2}{4\pi^2} T_B e^3_z .
\end{align}
where the cutoff for the Weyl spectrum is taken at $\Lambda^2 =  \left(\frac{c_{\perp}}{c_{\parallel}}\right)^2$, corresponding to Eq. \eqref{eq:Weyl_momenta} with $\frac{1}{2} \ll 1$. Remarkably {the LL results matches the more general torsional contribution} for the NY anomaly including curvature, as implied by the {anomalous} momentum non-conservation in the system as found in Ref. \onlinecite{Nissinen2019}. This result was found by matching the anomaly on emergent spacetime of background the chiral $p$-wave system to the corresponding BCS hydrodynamic result of the superfluid. In particular, including the effects of superflow leads to a spin-connection and curvature perpendicular to $\unitvec{l}$, as required by the Mermin-Ho relations \cite{MerminHo76}. 

In the chiral superfluid (or superconductor) the above result holds for both the linear quasirelativistic and the anisotropic Newton-Cartan spacetime, as defined by the tetrad \eqref{eq:3HeA_tetrad}. This simply follows from the fact that the cutoff for the validity of {both models} coincides with \eqref{eq:Weyl_momenta}. In this case, therefore, the anisotropic model {NC} is expected to require the same cutoff {as the linear model} since the system is probed also in the perpendicular direction. This morally happens since $\unitvec{l}=\unitvec{m}\times\unitvec{n}$, making the triad dependent \cite{MerminHo76, LiuCross79, Nissinen2019}. Strictly speaking in the LL-model we approximated $\unitvec{l} \approx \unitvec{z}$ which for the general non-trivial textures is given higher order corrections \cite{CombescotDombre86}.

\subsubsection{Axial current}

{On the other hand}, for the non-relativistic {anisotropic NC} model, however, we can also compute the anomalous vacuum current, corresponding to the anomalous superfluid momentum from the filled states below $p_F$ \cite{Volovik85}. {The global spectrum has correct form, valid also outside the vicinity of the Weyl points}. The anomalous momentum current is given by
\begin{align}
\vek{j}_{\rm anom,\parallel} = -2 \int^{p_F}_{0} dp^3 N_{\rm LL}(p_z) p_3  = -\frac{p_F^3}{6\pi^2} \unitvec{l}(\unitvec{l} \cdot \nabla \times \unitvec{l}) \label{eq:vacuum_current}
\end{align}
and even extending to $p_z=0$, there is no need for a cutoff. See Fig. \ref{fig:quadratic_spectrum}.

This is actually the correct hydrodynamic result for the (weak-coupling) BCS system \cite{VolovikMineev81, Volovik85, CombescotDombre86} to lowest order in gradients, since the final answer for the anomalous vacuum current is sensitive only to the $e_3 = \unitvec{l}$ direction, even in the presence of $\vek{v}_s$ (corresponding to curvature in the perpendicular plane). Upon taking the time-derivative of this momentum, the hydrodynamics of the system produce the covariant current implied by the Weyl anomaly.  If we assume, without any supporting arguments, that the curvature and torsion contribute to the current \eqref{eq:vacuum_current} as they enter the anomaly Eq. \eqref{eq:NYanomaly}, we get the same result if we apply the cutoff \eqref{eq:Weyl_momenta} as above, {even in the linear model}. We note that these findings are corroborated by the thermal contribution to the NY anomaly, as found in Ref. \cite{NissinenVolovik2019}. {The proper inclusion of curvature also ensures that states far away from the Fermi surface do not contribute to the currents}.

These considerations beyond the LL spectral flow {aside}, what we want to here emphasize is that the \eqref{eq:vacuum_current} current corresponds to the consistent anomaly, and can be derived from a corresponding Wess-Zumino terms {that should be generalized for torsional spacetimes \cite{Volovik1986c, Balatsky87, PeetersWaldron99, Landsteiner16, KurkovVassilevich18, Stone2019b, Copetti20}}. See especially \cite{Copetti20}, where the consistent and covariant anomalies are discussed in an anisotropic Lifshitz model, closely related to Eq. \eqref{eq:NC_fermion}. We leave the study of the consistent vacuum current from the perspective of gravitational anomalies with torsion for the future.

\section{Strained Weyl semimetals}\label{sec:WSM}
Semimetals with Weyl fermions arise in solid-state systems where the Fermi energy is tuned to a band-crossing in the Brillouin zone \cite{NielsenNinomiya83, WanEtAl11}. The tetrads arise universally via the {coefficients of the} linear expansion. In this case, the fermions are also charged leading to the possibility of the U(1) anomaly with electric fields \cite{NielsenNinomiya83}. In addition to the tetrads, {related} effective background (axial) fields {can be considered} with similar origin as in the chiral superconductor \cite{Volovik03} -- the (constant) shift of the Weyl node in momentum space {that leads} to the existence of the protected Fermi arc states \cite{Haldane14, Landsteiner16, GrushinEtAl16}. Here we would like to clarify {the related but physically distinct torsional contribution to anomalous transport} from the tetrads in the presence of elastic strains. In fact, due to the universal coupling of the tetrads to momentum \cite{ParrikarEtAl14, ShapourianEtAl15}, as in gravity, one expects that deformations of the (lattice) geometry would lead to effects that probe the Weyl fermions via the background tetrads. {This framework correctly takes into account the anomalous physics of the momentum dependent fields, see nevertheless \cite{ZhouEtAl13, SunWan14, Fujimoto16, PikulinEtAl16, GrushinEtAl16, HuangEtAl19, FerreirosEtAl19, Stone2019, HuangBo20}.} 

We start in a roundabout way, first discussing the low-energy Weyl Hamiltonian and then considering a lattice model for a realistic $T$-breaking material.

\subsection{Bloch-Weyl fermions in crystals}

The low-energy Bloch-Weyl Hamiltonian is of the form \cite{NielsenNinomiya83, WanEtAl11, ArmitageEtAl18}
\begin{align}
h_{\pm}(\vek{k}) &= \pm \sigma^a (k_a \mp k_{F,a}) + \hc \nonumber\\
&= \pm \frac{\sigma^a}{2} e^{i}_{a}(k_i \mp k_{F,i}) +\hc .
\end{align}
where now
\begin{align}
e^i_a = \frac{\doo H_{\rm TB}(\vek{k})}{\doo k^a}\bigg\vert_{\vek{k}_F} 
\end{align}
are simply the linear coefficients of the expansion of the underlying (tight-binding) Bloch Hamiltonian $H_{\rm TB}(\vek{k})$ near the Weyl nodes. Before we consider lattice deformations in this model, we remark on the interplay of the tetrads and momentum. The lattice momentum is \cite{ShapourianEtAl15}
\begin{align}
\hat{p}_a = \frac{\im}{2a} \sum_{\vek{x}}  c_{\vek{x}}^\dagger c_{\vek{x}+\unitvec{a}}- c_{\vek{x}+\unitvec{a}}^\dagger c_{\vek{x}} = \sum_{\vek{k}} \sin (k_a a) c^{\dagger}_{\vek{k}}c_{\vek{k}} .
\end{align}
Under non-trivial background fields, the Weyl system itself is anomalous under the lattice translation symmetry, $T_{3} = T_{\unitvec{z}}$, corresponding to the conservation of the lattice momentum $\hat{p}_3$,
\begin{align}
T_{\unitvec{z}}^{\dagger} c_{\pm \vek{k}} T_{\unitvec{z}} = e^{\pm\im a\vek{k}_{\rm F}}c_{\pm \vek{k}_F} \label{eq:lattice_rotation} 
\end{align}
which corresponds to an anomalous chiral rotation of the low-energy Weyl fermions at the $T$-breaking nodes $\pm \vek{k}_F$. Here $c^{\dagger}_{\vek{k}}$ creates the state corresponding to the lattice periodic Bloch state $\vert v_{\vek{k}}\rangle = \vert v_{\vek{k}+\vek{K}} \rangle$, with wave function
\begin{align}
\psi_{\vek{k}}(\vek{x}) = e^{\im \vek{k}\cdot \vek{x}}v_{\vek{k}}(\vek{x}).
\end{align}
In the presence of elastic deformations corresponding to torsion, i.e. phonons, the anomalous chiral symmetry corresponding to translations is manifested as the non-conservation of {(lattice)} momenta between the Weyl fermions and the background phonons \cite{Nissinen2019, Burkov20}, as found in superfluid $^3$He-A for the $p+ip$-wave paired Fermi-liquid \cite{Volovik03}. See also \cite{CortijoEtAl15, FerreirosEtAl19, NissinenVolovikPRR19, Copetti20}.

\subsection{Elastic deformations}
Now consider general lattice deformations. The original unstrained lattice momenta entering the Weyl {Hamiltonian} are represented as $k_a$ and the deformed lattice is given as $k_i = e^{\ a}_i k_a$ in the coordinate system of the laboratory, where $e^{\ a}_{i} \neq \delta^a_i$ to first order in the strains. These will couple as expected in the continuum model, as long as we take into account the lattice model properly, as we now recall following \cite{ShapourianEtAl15}. See also \cite{FerreirosEtAl19}. We have the continuum linear strain tensor,
\begin{align}
e^{\ a}_i = \delta^a_i + w^{\ a}_i &= \delta^a_{i}+\doo_i u^a \nonumber\\
e_{\ a}^i = \delta^i_a - w^i_{\ a} &= \delta_a^{i}-\doo_j u^b \delta_{ab} \delta^{ij} \label{eq:continuum}
\end{align}
where $u^a/a \ll 1$, {in terms of the lattice constant}. This means that $k_{F,a}$ is held fixed, whereas $k_{F,i}$ with $\delta k_{F,i} = w_i^{\ a} k_{F,a}$ is deformed ({in the laboratory coordinates}). This becomes on the lattice
\begin{align}
k_a \to k_a -w_{\ a}^i \frac{\sin k_i a}{a} \approx e_{\ a}^i k_i, \nonumber \\
k_i \to k_i + w^{\ a}_i \frac{\sin k_a a}{a} \approx e_{i}^{\ a}k_a . \label{eq:lattice}
\end{align}
where $w_{\ a}^i = \doo_j u^{b} \delta_{ab}\delta^{ij}$ is defined above and in the last approximation, the linear approximation for strain as well as $k_i a \ll 1$, close to the $\Gamma$-point, are used. In addition we assume that we work with low-frequencies corresponding to the acoustic phonons, below the Debye energy \cite{ShapourianEtAl15}.

\subsection{Lattice model}

In general, a model for a $T$-breaking Weyl semimetal consist of layered 2D Wilson fermions tuned to a zero energy crossing in three dimensions \cite{Volovik03, SukhachovEtAl18}. For a model of this kind pertaining to a real material, Ref. \cite{PikulinEtAl16} considered a time-reversal invariant $k\cdot p$ close to the $\Gamma$-point, where the the Weyl node itself will be at finite momentum corresponding to four momenta in the Brillouin zone, the minimum for $P$-breaking system. While the $k\cdot p$ model is realistic, it is more convenient to work with an explicit model with a lattice regularization that produces the same results. In terms of a tight-binding model, they considered
\begin{align}
H_{\rm lat}(\vek{k}) = \epsilon(\vek{k}) + \left(\begin{matrix} h_{\rm lat}(\vek{k}) \\ & -h_{\rm lat}(\vek{k}) \end{matrix}\right), \label{eq:H_latt}
\end{align}
where we focus on the time-reversal odd block $h_{\rm latt}(\vek{k})$ of the $T$-invariant model \cite{Volovik03, PikulinEtAl16, SukhachovEtAl18},
\begin{align}
h_{\rm lat}(\vek{k}) = t_z(M - \sum_{i=x,y,z} c_{i} \cos k_i a) \sigma^3 \\
+ (t_x \sin k_xa ) \sigma^1 + (t_y \sin k_ya) \sigma^2 . \nonumber
\end{align}
For $-1<\frac{M-c_x-c_y}{c_z}<1$ the model $h_{\rm lat}(\vek{k})$ has Weyl points at 
\begin{align}
\pm a\vek{k}_F  = (0,0,\pm \arccos \frac{M-c_x-c_y}{c_z}), 
\end{align}
otherwise it is gapped. The dimensionful tetrads are 
\begin{align}
e^i_a(\pm \vek{k}_{F}) = a(t_x, t_y, \pm t_zc_z \sin a k_{F,z})\delta^i_a. 
\end{align}
Inversion symmetry $P$ acts as $h_{\rm lat}(\vek{k}) \to \sigma^z h_{\rm latt}(-\vek{k}) \sigma^z$. For simplicity we set $c_z=1$, $c_{x,y} = c_{\perp}$, $t_{x,y} = t_{\perp}$ and assume uniaxial symmetry along $\unitvec{z}$ in the following.  We expect \eqref{eq:lattice} to hold for the Weyl semimetal model Eq. \eqref{eq:H_latt}, originating from the $k\cdot p$ model {close to the $\Gamma$-point}.

For this tetrad we can {moreover} ignore the difference of lattice and coordinate indices, with $u_{ij} = \frac{1}{2}(\doo_i u_j + \doo_j u_i) + O(u^2)$ the symmetric lattice strain. The strain induces the deformation considered in Ref. \cite{CortijoEtAl15} and \cite{PikulinEtAl16, GrushinEtAl16}
\begin{align}
\delta h_{\rm lat}(\vek{k}) =& - t_z \beta_{\rm el}u_{zz} \sigma^3 \cos ak_z \nonumber\\ 
&+ t_{\perp}\beta_{\rm el}(u_{xz} \sigma^1+u_{yz} \sigma^2) \sin ak_z
\end{align}
which gives
\begin{align}
\delta e^i_a = a t_z \beta_{\rm el} u_{ii} \delta_a^i \sin (k_Fa) + at_{\perp} \beta_{\rm el} \sum_{i' \neq i} u_{ii'}\delta^{i'}_a \cos (k_F a)
\end{align}
where $\beta_{\rm el}$ is the Gr\"unesein parameter. Restricting to a uniaxial strain corresponding to the axis of the Weyl node orientation, with the approximation that $ak_F\ll 1$,
\begin{align}
e_a^z \to  at_z (1+\beta_{\rm el}u_{zz})\delta_{a3} + a t_{\perp} \sum_{i=x,y} \beta_{\rm el}u_{zj}\delta_{a}^j, \nonumber \\
\delta e_3^z = at_z u_{zz}, \quad \delta e^z_1 = at_\perp u_{zx}, \quad \delta e^z_2 = a t_{\perp} u_{yz}. 
\end{align}
This has the (dimensionless) inverse tetrad, up to the neglected terms $O(u^2)$ in strains,
\begin{align}
e^1_i &= \unitvec{x}, \quad e^2_i = \unitvec{y}, \nonumber\\ 
e^3_i &= \unitvec{z}-\beta_{\rm el}\left(u_{zx},\left(\tfrac{t_z}{t_\perp}\right) u_{zy},\left(\tfrac{t_z}{t_\perp}\right)u_{zz}\right) .
\end{align}
This is what we expected, based on the corresponding universal continuum limit \eqref{eq:continuum} and the lattice substitution \eqref{eq:lattice} coupling to geometry, apart from the (non-universal) couplings $\beta_{\rm el}$, $\left(\tfrac{t_z}{t_\perp}\right) $ between the phonons and electrons of the lattice model \cite{ShapourianEtAl15}.
Now in the presence of non-homogenous strain vector $e^3_z $ depending coordinates and time, torsion $T^3_{\mu\nu}$ {and spectral flow} will arise. The Landau level arguments of Sec. \ref{sec:torsional_LLs} and \ref{sec:chiral} apply for a torsional magnetic field from $u_{zx,zy}(x,y)$ (in the ``symmetric gauge") and an adiabatic electric field from $u_{zz}(t)$, as in \cite{PikulinEtAl16, GrushinEtAl16}.

\subsection{Torsional density of states in anomalous transport}

Armed with the geometric background fields corresponding to torsional (magnetic field), we can consider the anomaly resulting from the chiral rotation \eqref{eq:lattice_rotation}. The linear Weyl model is valid up to the approximation
\begin{align}
t_z(M - \sum_{i=x,y,z} c_{i} \cos k_i a) & \\
\approx \frac{t_za^2}{2} \bigg[c_{\perp}(k_x^2+k_y^2) & +(k_z \mp k_F)^2\bigg] \nonumber\\
\approx t_z a e_3^i(k_i-&k_{F,i}) = (t_za \sin k_Fa)q_{z}
\end{align}
which is simply restricted by the ignored terms of the remainder in the expansion. Apart from the trivial $q_z \ll k_F \ll 1/a$, also
\begin{align}
c_{x} \cos q_x a + c_{y}\cos q_ya  \nonumber
\approx& \frac{c_{\perp} a^2}{2} (q_x^2 + q_y^2) =  \frac{c_{\perp} a^2}{2}q_\perp^2\\
\ll& \frac{t_x}{t_z}a q_x + \frac{t_y}{t_z}a q_y = \frac{t_{\perp}}{t_z}a q_{\perp}
\end{align}
leading to the constraint $q_{\perp} \ll \frac{2 t_\perp }{c_\perp a t_z}$, meaning
\begin{align}
E_{\rm Weyl} \ll  \frac{t_\perp^2}{c_{\perp} t_z},
\end{align}
for the perpendicular direction. We are working in the units where $-1<M-2c_{\perp}<1$ and $\cos k_Fa = M-2c_{\perp} \approx 1$. For the effects of any torsional anomaly from {magnetic strain}, we can just evaluate the chiral densities at the nodes,
\begin{align}
n_{\pm}(\Lambda) = ej^0_{\pm} = \int_{\pm k_F(1-\frac{\Lambda^2}{2})}^{\mp k_F(1+\frac{\Lambda^2}{2})} dk^3 N_{\rm LL}(k_z) \nonumber \\
=\mp \frac{k_F^2 \Lambda^2}{4\pi^2}\beta_{\rm el}\left(\tfrac{t_z}{t_\perp}\right)T_B e^3_z . 
\end{align}
It is interesting to recall that for the chiral superfluid, while strictly it must be that $\Lambda^2 \ll 1$ since $q_{z} \ll k_F$, we found that the cutoff was parametrically high ``$\frac{1}{2} \ll 1$" in terms {of the validity of the Weyl description}. There however, due to the orthonormal triad, also the perpendicular direction couples to the transport, with the cutoff Eq. \eqref{eq:Weyl_momenta} which in real $^3$He-A is actually $\sim 10^{-6} p_F$. 

For the semimetal, the case where $q_z \sim \frac{t_{\perp }}{t_{z} \sin k_Fa}q_{\perp} \ll k_F$ {arises when} assuming that we isotropically couple to the perpendicular directions for general {strain} field configurations. Plugging in real parameters, we expect that for e.g. Cd$_3$As$_2$, $t_\perp \sim t_z \sin k_Fa$ \cite{PikulinEtAl16}.  {Another option would be to consider the Newton-Cartan model with quadratic spectrum $M-2c_{\perp}-\cos k_za$ along the Weyl node direction with {uniaxial strain} only, with the constraint $q_z \ll k_F$}. The same model with different parameter also applies for the Dirac semimetal Na$_3$Bi \cite{PikulinEtAl16} and references therein.

Independent of whether one has a torsional electric field $\doo_t e^3_z \neq 0$ or an electric field $E^z$ driving the spectral flow, as in Fig. \ref{fig:B5E} and \ref{fig:B5E5}, this will lead to the suppression of the density proportional to $\Lambda^2$, corresponding to the validity of the linear Weyl approximation, in the anomalous transport, as compared to the Fermi wavevector $k_F$ and {the pseudo gauge field in momentum space \cite{PikulinEtAl16, GrushinEtAl16}}. We note that this reduction of {anomalous axial density} is simply due to the momentum dependent density of states. {This, as we have explained, naturally follows from the tetrads and torsion coupling to momenta and should be contrasted with a U(1) gauge field and constant density of states, as dictated by the universal minimal coupling and the topology of U(1) gauge fields}.

\section{Thermal effects}\label{sec:thermal}
Finally we briefly recall and discuss thermal contributions to the torsional anomaly. There are two possible effects: i) the small but finite temperature enters the NY anomaly as the scale of thermal fluctuations in momentum space. These are analyzed in \cite{NissinenVolovik2019, NissinenVolovik19b, Stone2019} ii) There is a {related} finite thermal gradient in the system and one computes the thermal response via Luttinger's fictitious gravitational field \cite{Luttinger64}. We note that non-zero time-like torsion for the Luttinger spacetime implies the non-single valued time coordinate in the fictitious gravitational field \cite{BradlynRead15}. See also \cite{Stone12, GromovAbanov15, Sekine16, RyuEtAl17, ChernodubEtAl18, KobayashiEtAl18}.

Here we focus on the effects of a thermal gradient, the currents induced can be computed by coupling the system to fictitious spacetime metric, following Luttinger \cite{Luttinger64}. Specifically, we assume a thermal gradient
\begin{align}
\nabla \sigma = -\frac{1}{T}\nabla T 
\end{align}
which is equivalent to a weak gravitational potential $g_{00} = 1+2\sigma$ in the system. The perturbation $\delta g_{00}$ couples to the Hamiltonian (energy current) $T^{00}$. In units where the velocity of propagation is $v=1$, the metric is
\begin{align}
ds^2 &= e^{+2\sigma}dt - \delta_{ij}dx^i dx^j \\
&\approx (1+2\sigma)dt^2 - \delta_{ij}dx^i dx^j
\end{align}
from which the linear response to the thermal gradient $\sigma$ can be calculated \cite{Luttinger64}. This can be generalized to a metric
\begin{align}
ds^2 = e^{2\sigma}(dt+e^{-\sigma}N_i dx^i)^2 - \delta_{ij}dx^i dx^j \\
= e^0_{\mu}e^{0}_{\nu} dx^{\mu}dx^{\nu} - \delta_{ij} d x^i dx^j,
\end{align}
now with a small gravimagnetic potential \cite{Volovik03, RyuEtAl17}
\begin{align}
A_{\mu}^{\rm g} = (e^{\sigma},N_i) \approx (1+\sigma, N_i) \equiv e^0_{\mu},
\end{align}
where $N_i$ describes a velocity field in the units where $v=1$. The gravitational thermal potential \cite{Volovik03, RyuEtAl17, KhaidukovZubkov2018}
\begin{align}
-\frac{1}{T}\nabla T = \nabla \sigma - \doo_t N_i. \label{eq:gravimagnetic}
\end{align}
whence 
\begin{align}
e^0_{\mu} &= (e^{\sigma}, N_i), \quad e^a_\mu =\delta^{a}_{\mu}, \quad a=1,2,3 \\
e^{\mu}_0 &= (e^{-\sigma},0), \quad e^{\mu}_a = (e^{-\sigma}N_i,\delta^{i}_{a}), \quad a=1,2,3. 
\end{align}
In this case Eq. \eqref{eq:gravimagnetic} becomes 
\begin{align}
-\frac{1}{T}\nabla T = \nabla \sigma - \doo_t N_i = \doo_{i}e^{0}_{t} - \doo_{t}e^0_{i} = T^{0}_{i t}
\end{align}
where $T^0_{\mu\nu}= \doo_{\mu}e^0_{\nu}-\doo_{\nu}e^0_{\mu}$ is the temporal torsion, assuming zero temporal spin-connection $\omega^0_{\mu b} \equiv 0$. It is expected then, that one would have possibility for anomalous transport in terms of the combination of thermal gradient and vorticity $T^0_{ij} = \doo_i N_j -\doo_j N_j$ in the velocity field $N_i(x)$, as in the chiral vortical (and magnetic) effect \cite{KhaidukovZubkov2018, ImakiYamamoto19}.

Now similarly as we expect momentum density at the Weyl node $(P^{\mu})_{\rm node} = \Pi^{t\mu} = p_F e_3^{i}\delta_{i}^\mu e j^{0}_5$\cite{Nissinen2019} for the Weyl systems at finite $p_{Wa}=p_F\delta_{3a}$, or since $T^{0\mu} = e e^{\mu}_a T^{t a}$,
\begin{align}
e \Pi^{t 3}= \frac{p_F^3 \Lambda^2}{16\pi^2} e^3_\mu e_3^i \delta_i^{\mu} \epsilon^{0\nu\lambda\rho} e^3_{\nu} T^3_{\lambda\rho}
\end{align}
we expect an energy density of the form
\begin{align}
J^{t}_{\epsilon} = eT^{t}_{\ 0} = p_F e  j^0_5= \frac{p_F T^2}{12v^2} \epsilon^{tijk} e_{i}^0 T^0_{jk} 
\end{align}
where $T^{\mu}_{\ a} \equiv \frac{1}{e}\frac{\delta S}{\delta e^a_\mu}$. The anomaly of this current would be proportional to $T\nabla T$, and is indeed reminiscent of the chiral vortical effect \cite{GromovAbanov15, KhaidukovZubkov2018}. We can also expect mixed terms, in the sense that there should be a corresponding energy current from \emph{both} the momentum density and thermal current, $\doo_t e^i_3 \neq 0$, at the node 
\begin{align}
J^i_{\epsilon} = e T^i_{\ 0} =  \frac{p_F T^2}{6v^2} \epsilon^{0ijk} e^3_j \times T^0_{0k} + \frac{p_F T^2}{12v^2} \epsilon^{0ijk} e_t^0 T^{3}_{jk} ,
\end{align}
these ``mixed" contributions to the currents were identified and discussed in Ref. \cite{LiangOjanen19b}.

The message we want to convey here is that one can indeed expect anisotropic and ``mixed" contributions to the torsional anomalies, in the sense that the Lorentz invariant $\Lambda^2\eta_{ab} \to \Lambda_a \Lambda_b$ a generalized anisotropic tensor, in various condensed matter systems depending on the symmetries, perturbations and cutoffs. We leave the detailed discussion of such thermal gravitational contributions for the future, see however \cite{Stone2019, LiangOjanen19b} and the general discussion in \cite{NissinenVolovik2019}.
 
\section{On the relation of emergent torsion and pseudo gauge fields} \label{sec:comparison}

Here we summarize our findings in relation to earlier literature, where the momentum space field corresponding to the shift of the node is often considered as an axial gauge field \cite{Volovik85, Volovik03, CortijoEtAl15, Fujimoto16, PikulinEtAl16, GrushinEtAl16, SukhachovEtAl17, HuangEtAl19, FerreirosEtAl19, IlanEtAl19}. We note that torsion can be shown to enter as an axial gauge field constructed from the totally antisymmetric torsion $\gamma^5S^{\mu} =\epsilon^{\mu\nu\lambda\rho}T_{\nu\lambda\rho}$ \cite{ChandiaZanelli97, Soo99} coupling to the momentum. This is essentially what we found in Secs. \ref{sec:torsional_LLs} and \ref{sec:chiral} with the momentum space dependent LL density of states. {The LL calculation and anomaly itself should be performed by taking this momentum dependence into account, as we have done here}. 

How are tetrads with torsion otherwise different from the momentum gauge field? The symmetries corresponding to the tetrads are translations which for finite node momenta, {requisite for condensed matter Weyl fermions}, corresponds to the anomalous chiral symmetry. There is no local gauge symmetry corresponding to the Berry curvature in momentum space. On the other hand, the geometric formulation is suited for such translation symmetries and reveals the background geometry of the spacetime emerging from the node \cite{Horava05}. The overall geometry can made consistent with the non-relativistic symmetries away from the Weyl node for a finite momentum range. For the anomalous axial density and anomaly, this leads to the parametric suppression compared to U(1) anomaly and the UV-scale $p_W$. The phenomenological implications of this are significant, even without the theoretical recourse to the emergent geometry.

We also note that Ref. \cite{FerreirosEtAl19} discusses torsion (and the conservation of momentum) in strained semimetals in terms of a model with both the axial gauge field from the node and the tetrad with elastic deformations. While such a ``splitting" between low-energy and high-energy momenta is in principle allowed, it makes the consideration of the momentum dependent anomalies more involved, with the danger of double counting. The momentum anomaly (without EM gauge fields) should be proportional  $k_W \doo_{\mu}(e j^{\mu}_5)$, as found in \cite{Nissinen2019}. 

The original paper \cite{ShapourianEtAl15} for {elastic} deformations takes an explicitly geometrical view point which nicely connects with the strain induced tetrad formalism proposed here. In the simplest possible terms, we start with the Weyl (or Dirac) Hamiltonian in flat space with the small deformation $e^i_a = \delta^i_a+\delta e_a^i$,
\begin{align}
H_{+} = \sigma^a(\hat{k}_a - k_{Wa}) &\to \frac{\sigma^a}{2} e^i_a (\hat{k}_i - k_{Wi}) + \hc \nonumber\\
&=  \frac{\sigma^a}{2} (e_a^i k_i - k_{Wa}) + \hc. \\
&\approx \frac{\sigma^a}{2} ([\delta_a^i + \delta e^i_a] q_i + k_W\delta e^i_a) + \hc \nonumber
\end{align}
where now $k_W \delta e^i_a =-k_W \delta e^a_i$ is the momentum space gauge field in the Hamiltonian with (almost) constant tetrads \cite{Volovik85, BalatskiiEtAl86, ShapourianEtAl15, PikulinEtAl16, GrushinEtAl16, FerreirosEtAl19}. The right-hand side is the Hamiltonian in coordinate (or laboratory) space, which is the one we have experimental access to, and is deformed with respect to the orthogonal frame of $k_a$. We see that the momentum $\hat{k}_i$ couples to $e^{i}_a$, as expected, and the shift is essentially constant in the Hamiltonian, in the sense that $k_{Fa}$ is constant corresponding to the undeformed case, irrespective of the deformation. At the same time, the laboratory value changes though as $k_{Fi} = e^a_i k_{Fa}$.  In the examples we considered, in the chiral superfluid and superconductor we explicitly have that $k_{F,i}=p_F e^3_i$, giving $k_{Fa} = p_F\delta^3_a$. Similarly, for the strained semimetal we consider the originally unstrained lattice Fermi wave vector $k_{Fa}(x) \to k'_{Fa}(x+u) \approx k_{Fa}(x) + \doo_i u^a k_{Fa}(x) \equiv e_i^a k_{Fa}$ under strain $x' = x+u$, giving Eq. \eqref{eq:continuum} as expected.

What this means more generally is that $\nabla k_{Fa}=0$, in terms of the connection corresponding to the emergent spacetime, as discussed in Sec. \ref{sec:spacetimes}. In fact this is one of the requirements for the consistent assignment of the low-energy geometry.  On the other hand, all the torsional spacetimes we considered are in some sense abelian (or gravitoelectromagnetic) since the relevant fields can be identified as an abelian gauge fields in momentum space, amounting to what was called ``minimal coupling" trick in \cite{ParrikarEtAl14,ShapourianEtAl15}. In this case however, the gravitational character comes still evident in the momentum dependent charge and density of LLs, as expected for gravitational response, coupling to momenta and energy densities {including thermal effects}. 

\section{Conclusions and outlook}\label{sec:conclusions}

In this paper, we have argued for the emergence of non-zero torsional anomalies in Weyl (and Dirac) systems with simple Landau level arguments. In particular, we were motivated by the possibility of non-zero torsional Nieh-Yan anomalies in condensed matter systems with an explicit cutoff and the lack of relativistic Lorentz symmetries. For the anomaly, the spectral flow in the presence of torsion clearly renders non-zero results for Weyl nodes at finite momentum. Although obtained with specific simple field configurations corresponding to the torsion with Landau level spectra, they are expected to generalize covariantly in terms of the relevant spatial symmetries of the system. We discussed two idealized spacetimes related to the {symmetries}, the linear Riemann-Cartan and the anisotropic Newton-Cartan spacetime with quadratic dispersion. 

{We also briefly discussed the thermal torsion via Luttinger's fictitious spacetime, since we can expect mixed anomalies already from the inclusion of thermal gradients. This connects to gravitational anomalies and transport in general \cite{NissinenVolovik2019}. The recent results on universal anomaly coefficients in linear response thermal transport related to gravitational anomalies  \cite{Landsteiner11, LoganayagamSurowka12, JensenEtAl13, Landsteiner2014, LucasEtAl2016, StoneKim18} are related. From the non-universal torsional anomaly, via e.g. the momentum dependent LL density of states, the expected gravitational anomaly polynomials at finite temperature arise already at the level of linear response from the universality of IR thermal fluctuations \cite{NissinenVolovik2019}.} Moreover, we expect that the emergent tetrads with coordinate dependence arise rather generally in any Weyl system, making sense of evaluating the linear response to these, even in flat space. 

We {clarified} the relation between momentum space pseudo gauge fields and the emergent tetrads. It is important to realize that the spectral {or Hamiltonian} correspondence between torsion and U(1) magnetic fields, e.g. in a Landau level problem, is not yet enough for the anomalies to match in general. The simple LL spectral flow argument is enough to identify the non-universal cutoff appearing in the NY anomaly term. The message is that {low-energy tetrads and geometry} couple to the momentum in a universal way, even in lattice models with some caveats \cite{ShapourianEtAl15, CortijoEtAl15}, due to the non-universal coupling of the lattice phonons and fermions {as compared to pure continuum}. The UV scales appearing in the termination of anomalous chiral transport from such emergent fields, related to the Fermi-point momentum $p_W$ and the regime of validity of the effective Weyl/Dirac description, are naturally understood from the geometric perspective. In the presence of both independent U(1) fields and momentum space tetrads we should also expect many mixed terms, as studied e.g. in \cite{KubotaEtAl01, ParrikarEtAl14}. The mixed {torsional anomalies should also be carefully reconsidered with regards to finite node momentum, where we again expect differences to relativistic fermions}. {On this note} our results for the anomaly at finite momentum are in contrast to \cite{HuangBo20}, where a model with torsion is compared to a relativistic model at $p=0$ with pseudo gauge fields without consideration of node momentum coupling to the torsion or the cutoff of the quasirelativistic dispersion.

More formally, what we did amounts to applying the $K$-theory theorem of Horava \cite{Horava05} to the geometry of specific Weyl nodes in three dimensions, by keeping track of the UV symmetries and scales in the problem for the precise the form of the emergent geometry and fields coupling to the quasiparticles. The topology only guarantees the effectively Dirac like spectrum, with everything else depending on the microscopics. 

Many interesting avenues remain in the geometric description of topological condensed matter systems with gapless fermions, {including also nodal line systems \cite{NissinenVolovik2018, Schnyder20}}. It would be extremely interesting to study the gravitational anomalies in Weyl and Dirac systems from the global symmetry perspective with many nodes Weyl, taking into account the relevant space group symmetries \cite{CortijoEtAl15, Manes12, JuricicEtAl12, SlagerEtAl13, RaoBradlyn20}. More generally, the appearance of low-energy quasirelativistic fermions {with exotic geometric backgrounds within feasible experimental reach is expected to give} more insight also to the physics of relativistic gravitational anomalies with torsion \cite{ChandiaZanelli97}, although the symmetries and status of the background fields {are} dramatically different. 

\emph{Acknowledgements. ---}  We thank Z.-M. Huang for correspondence on his work, T. Ojanen and P.O. Sukhachov for discussions. Finally we especially thank G.E. Volovik for discussions, support and collaborations on related subjects. This work has been supported by the European Research Council (ERC) under the European Union's Horizon 2020 research and innovation programme (Grant Agreement no. 694248).

\appendix
\section{Review of the chiral anomaly and the spectral flow argument} \label{sec:appendix_EM}
\subsection{Weyl fermions in vector and axial U(1) fields}
The simplest way to argue for the axial anomaly in condensed matter systems is the spectral flow argument, utilizing Landau level spectrum in 3+1d \cite{NielsenNinomiya83, Volovik85}. To that end, one envisages a Hamiltonian of the form
\begin{align}
H_{\rm R,L} = \pm \sigma^i (\im \doo_i -q A_{i; \rm R,L})
\end{align}
where $A_i$ is some U(1) gauge field with charge $q$, i.e. the charges in the system are quantized in terms of $q$. Note that we still assume that the vector and axial combinations could still be both non-zero, 
\begin{align}
A_{\mu}=\frac{1}{2}(A_R+A_L)_{\mu}, \quad A_{5,\mu} = \frac{1}{2}(A_{\rm R}-A_{\rm L})_{\mu}.
\end{align}
Under these gauge fields, the chiral fermions always experience the chiral anomaly, where the classical conservation laws corresponding to these fields are broken.

The are summarized by the anomaly equations \cite{Landsteiner16, SukhachovEtAl17, SukhachovEtAl18, SukhachovEtAl18b, IlanEtAl19}
\begin{align}
\doo_{\mu} j^\mu &= \frac{q^2}{8\pi^2} \epsilon^{\mu\nu\lambda\rho} F_{\mu\nu} F_{5\lambda \rho} \label{eq:U1_anomaly_eqs}\\
\doo_{\mu} j^\mu_5 &= \frac{q^2}{16\pi^2} \epsilon^{\mu\nu\lambda\rho}( F_{\mu\nu} F_{\lambda \rho} + F_{5 \mu\nu} F_{5 \lambda \rho} ). \nonumber
\end{align}

Naturally, if the particles couple to both axial and vector gauge fields, countercurrents must be introduced to the system to conserve the total particle number, as required by conservation of charge. This is done by adding Bardeen-Zumino counterterms to the effective action \cite{Landsteiner16}, 
\begin{align}
    \Gamma[A,A_5] \to \Gamma[A,A_5] - \int \frac{d^4x}{12\pi^2} \levic F_{\mu\nu}A_\rho A^5_\sigma.
\end{align}
This is the Bardeen counter term, and it introduces the countercurrents
\begin{align}
        \delta j^\mu & = \frac{1}{12\pi^2}\levic (2F_{\rho\nu} A_\sigma^5 + F^5_{\rho\sigma} A_\nu) \nonumber\\
  \delta j^\mu_5 & = \frac{1}{12\pi^2}\levic F_{\nu \rho}A_\sigma.
\end{align}
These modify the anomaly in Eq. \eqref{eq:U1_anomaly_eqs} so that \(j^\mu\) is conserved: 
\begin{align}
    \partial_\mu j^\mu & = 0 \nonumber\\
    \partial_\mu  j_5^\mu & = \frac{1}{16\pi^2}\levic  \left( 3F_{\mu\nu}F_{\rho\sigma} + F_{\mu\nu}^5F_{\rho\sigma}^5\right).
\end{align}
Please see Refs. \cite{Landsteiner16} and \cite{SukhachovEtAl18, SukhachovEtAl18b} for more discussion.

\subsubsection{Landau levels}
We consider the minimally coupled Weyl Hamiltonian with vector potential \(\vek{A}=(-By, 0, 0)\),
\begin{align}\label{eq:Hmag}
    H_\ch &  = \ch\sigma^i(\hat{p}_i-qA_i) \\
    &= \ch\begin{bmatrix}\hat{p}_z && \khat_x+qBy - i\khat_y\\ \khat_x+qBy + i\khat_y && -\hat{p}_z \end{bmatrix},
\end{align}
where \(\ch = \pm1\) denotes the chirality fo the fermion. With an eigenstate ansatz \(\psi = e^{i(p_zz+p_xx)}\phi \) the eigenvalue problem becomes 
\begin{align}
    H_\ch\psi&  = \ch\begin{bmatrix}p_z && p_x+qBy - i\khat_y\\ p_x+qBy + i\khat_y && -p_z \end{bmatrix}\psi.
\end{align}
For \(qB > 0\) the off-diagonals can be identified as raising and lowering operators for a harmonic oscillator in the y-direction (displaced by \(p_x\)),
\begin{equation}
    \begin{cases}
    \hat{a} = \sqrt{2qB}^{-1}\left[qBy+p_x + i \hat{p}_y \right] \\
    \hat{a}^\dagger = \sqrt{2qB}^{-1}\left[qBy+p_x - i \hat{p}_y \right],
    \end{cases}
\end{equation}
which satisfy the properties $\{\hat{a},\hat{a}^\dagger\} = 1$, \(\hat{a}\phi_n = \sqrt{n}\phi_{n-1}\), and $\hat{a}^\dagger\phi_n = \sqrt{n+1}\phi_{n+1}$ for eigenstates of the harmonic oscillator $\phi_n$. The eigenvalue equation becomes
\begin{align}
    H_\ch\psi&  = \ch\begin{bmatrix}p_z && \sqrt{2qB}\hat{a}^\dagger\\ \sqrt{2qB}\hat{a} && -p_z \end{bmatrix}\psi.
\end{align}
The energy eigenvalues are obtained from considering the squared Hamiltonian operator: 
\begin{equation}
    \begin{split}
        H_\ch^2 & = (\hat{\vek{p}} - q\vek{A})^2 - q\boldsymbol{\sigma}\cdot \vek{B} \\
        &= \khat_y^2 + p_z^2 + (p_x + qBy)^2 - qB\sigma_3
     \end{split}
\end{equation}
whence
\begin{align}
    E^2 &= p_z^2 + 2|qB|(n+1) - qB\sigma_3, \nonumber\\ 
    E &= \pm\sqrt{p_z^2 + 2|qB| n}, \quad n\geq 0. 
\end{align}
Looking now at the action of the ladder operators on components of the eigenstates \(\psi\), they must of the form 
\begin{equation}
    \psi =e^{i(p_zz+p_xx)}\begin{bmatrix}\phi_n \\ C_n\phi_{n-1}\end{bmatrix} 
\end{equation}
where \(\phi_n\) are eigenstates of the harmonic oscillator, \(\phi_{n-1}=0\) and \(C_n\) is a factor determined from the eigenvalue equation to be \(C_n = \dfrac{\sqrt{2qBn}}{\pm E+p_z}\) for \(n \neq 0\). The \(n = 0\) state is "half" occupied, since
\begin{align*}
    \psi = e^{i(p_zz+p_xx)}\phi_0\begin{bmatrix}1 \\ 0\end{bmatrix} 
\end{align*}
with chiral dispersion relation \(E = p_z\) for \(H_+\) and \(E = -p_z\) for \(H_-\), after the elimination of the trivial zero modes \(H_\ch\Psi \).

For \(qB < 0\) the spectrum is  the same but the eigenstates are now 
\begin{align}
    \psi_n &= e^{i(p_zz+p_xx)}\begin{bmatrix} D\phi_{n-1} \\ \phi_n \end{bmatrix}, \quad n \geq 1, \\ 
    \psi_0 &= e^{i(xp_x +zp_z)}\phi_0\begin{bmatrix}0 \\ 1\end{bmatrix}
\end{align}
where \(D = \dfrac{\sqrt{p_z \mp E}}{2|qB|n}\). The zeroeth Landau level dispersion relation is \(E = -p_z\) for \(H_+\) and \(E = +p_z\) for \(H_-\).

In summary: 
\begin{align}
\label{eq:relEMspectrum}
    E = \begin{cases}\pm \sqrt{p_z^2+2|qB|n}, \quad n\geq1 \\ \text{sgn}(qB\ch)p_z, \quad n = 0. \end{cases}
\end{align}
The degeneracy of each state can be determined from containing the system within a finite volume \(L_xL_yL_z\) and requiring the center of the harmonic oscillator be within it:
\begin{equation}
    0 \leq \frac{p_x}{|qB|} \leq L_y.
\end{equation}
The x-direction is free and is therefore quantized as \(p_x = n\frac{2\pi}{L_x}\) with \(n \in \mathbb{N}\). The z-direction is similarly quantized in units of \(\Delta p_z = \frac{2\pi}{L_z}\), so the number of states in the xy-plane per \(\Delta p_z\) is
\begin{equation}
\label{eq:dos}
    n = \frac{|qB|}{4\pi^2}L_xL_yL_z.
\end{equation}

\subsubsection{Spectral flow}
When an electric field parallel to $\vek{B}$ is turned on adiabatically, for example as \(\vek{A}=(-By, 0, -E_zt)\), the states flow in the spectrum according to Lorentz's law as \(\dot{k}_z = qE_z\).
The unpaired LLL chiral modes flow to specific direction, whereas the higher LLs cancel. The states consequently move in or out of the vacuum depending on their chirality as
\begin{align}
\label{eq:chiralcurrent}
   \part j_\ch^0 &= \text{sgn}(q\chi)\frac{q^2EB}{4\pi^2}  \nonumber\\
   &= -\sgn(q\ch)\frac{q^2}{32\pi^2}\levic F_{\mu\nu}F_{\rho\sigma} .
\end{align}

We need to generalize \eqref{eq:chiralcurrent} from Minkowski spacetime (with metric signature \(+---\)) to a general spacetime with or without torsion. The Landau level calculation genralizes to a non-trivial metric and coordinate dependent tetrads, when we work in the coordinate space $\tilde{x}^i \equiv e^i_a x^a $, $e d\tilde{V} =  dV$, where $e=\det e^a_i$, compared to the local Minkowski space. The invariant density of states and fields to be (for \(e^0_\mu = \delta^0_\mu\))
\begin{align}
    \frac{dN}{d V} dV &= \frac{|q B|}{4\pi^2} dV \\
    &=  \frac{d\Tilde{V}}{dV}\frac{dN}{d\Tilde{V}} e d\Tilde{V}  =  \frac{\vert q \Tilde{B} \vert}{4\pi^2} e d\Tilde{V},
\end{align}
which we need to use when we do not want the (scaling of the) tetrads to affect the physical density or flux we are interested in. With \(\dot{k_z} = -E_z\) the anomaly becomes
\begin{equation}
\frac{1}{e} \part (e j_0^\ch) = \frac{1}{e} \frac{\ch}{32\pi^2}\levic F_{\mu\nu}F_{\rho\sigma},
\end{equation}
which matches \eqref{eq:U1_anomaly_eqs} after covariantly generalized to a non-trivial metric. 

\subsubsection*{Inclusion of axial fields}

\begin{center}
    \begin{figure*}
    \centering
        \includegraphics[width=220pt]{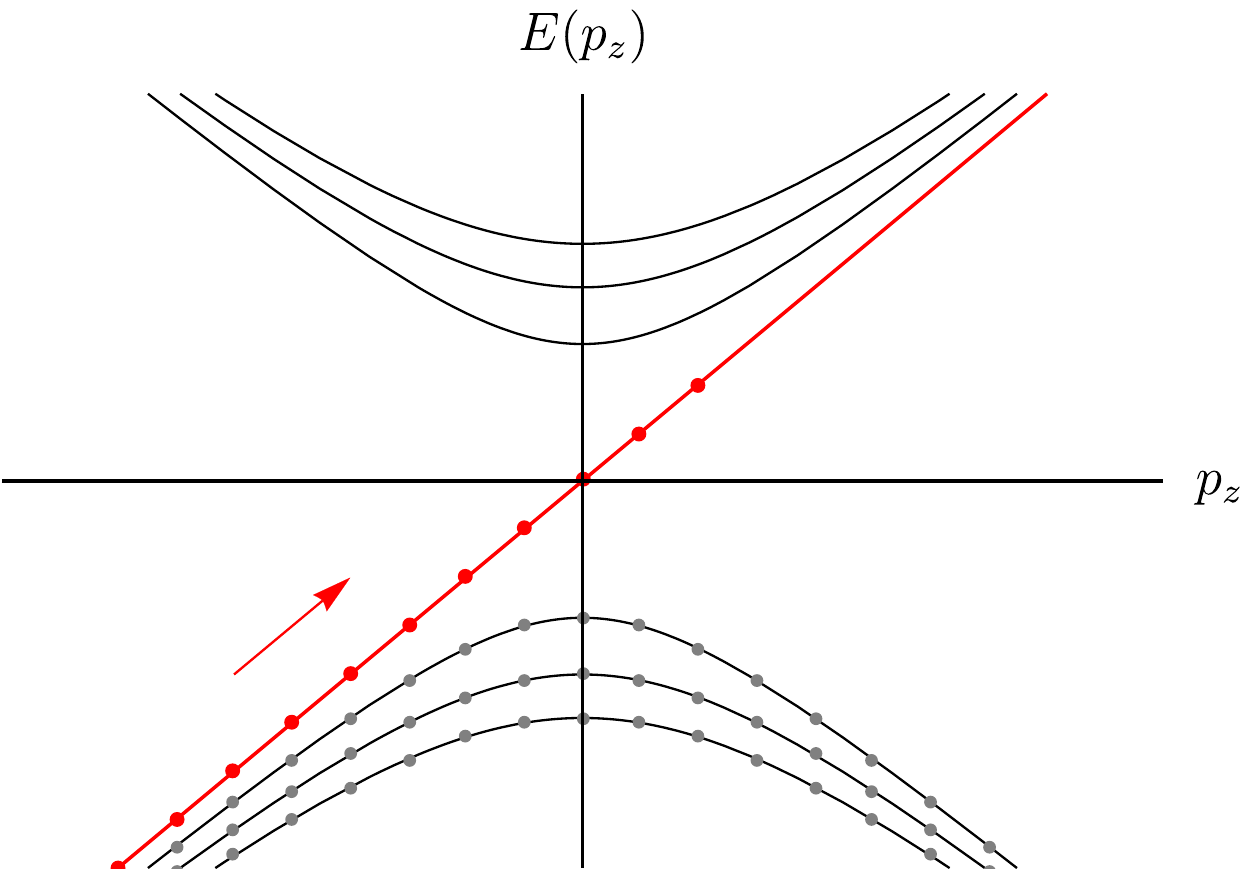}\includegraphics[width=220pt]{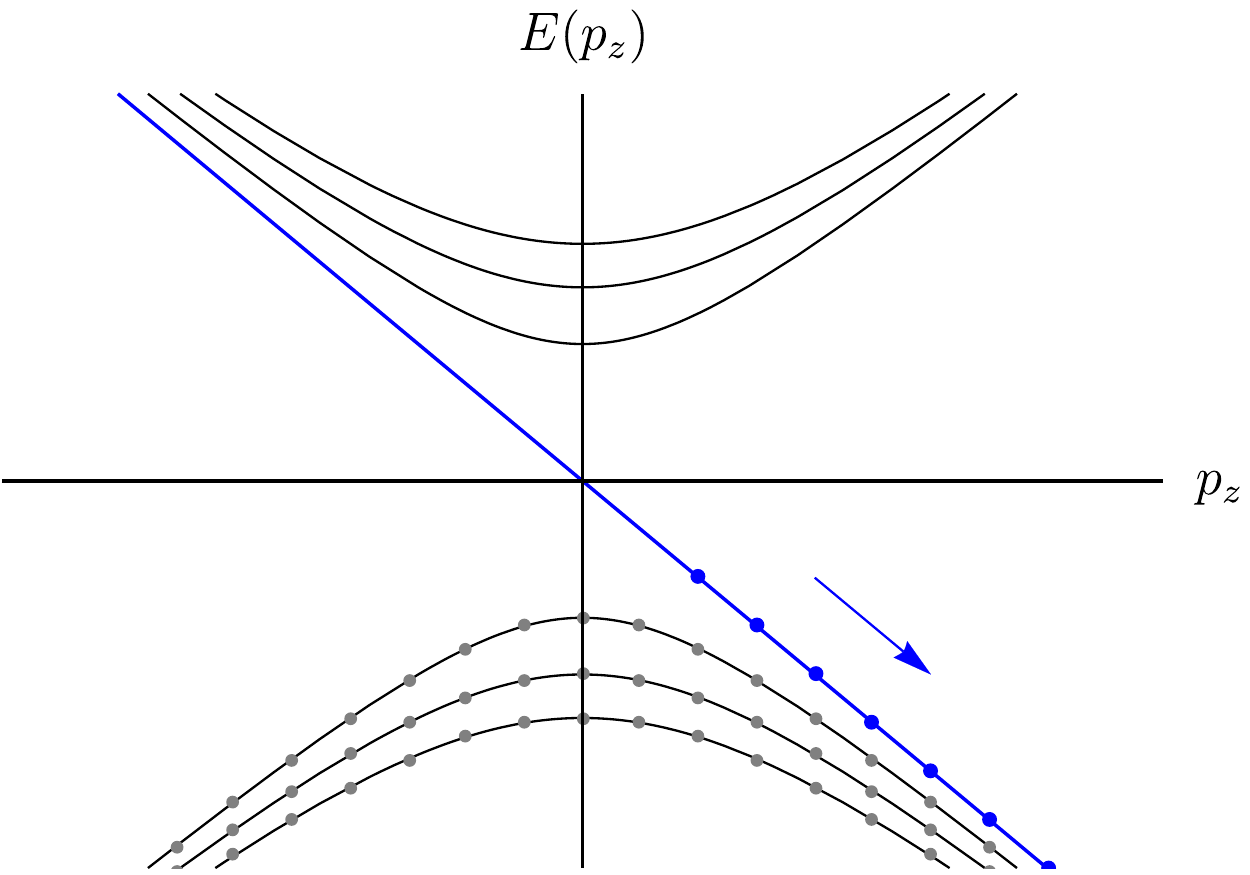}
        \caption{Dispersion and spectral flow of right-handed (red) and left-handed (blue) particles (\(q>0\)) under parallel $\vek{B},\vek{E}$.}
        \label{fig:BE}
        \includegraphics[width=220pt]{Kuvaajat/EM_kaikki_RH.pdf}\includegraphics[width=220pt]{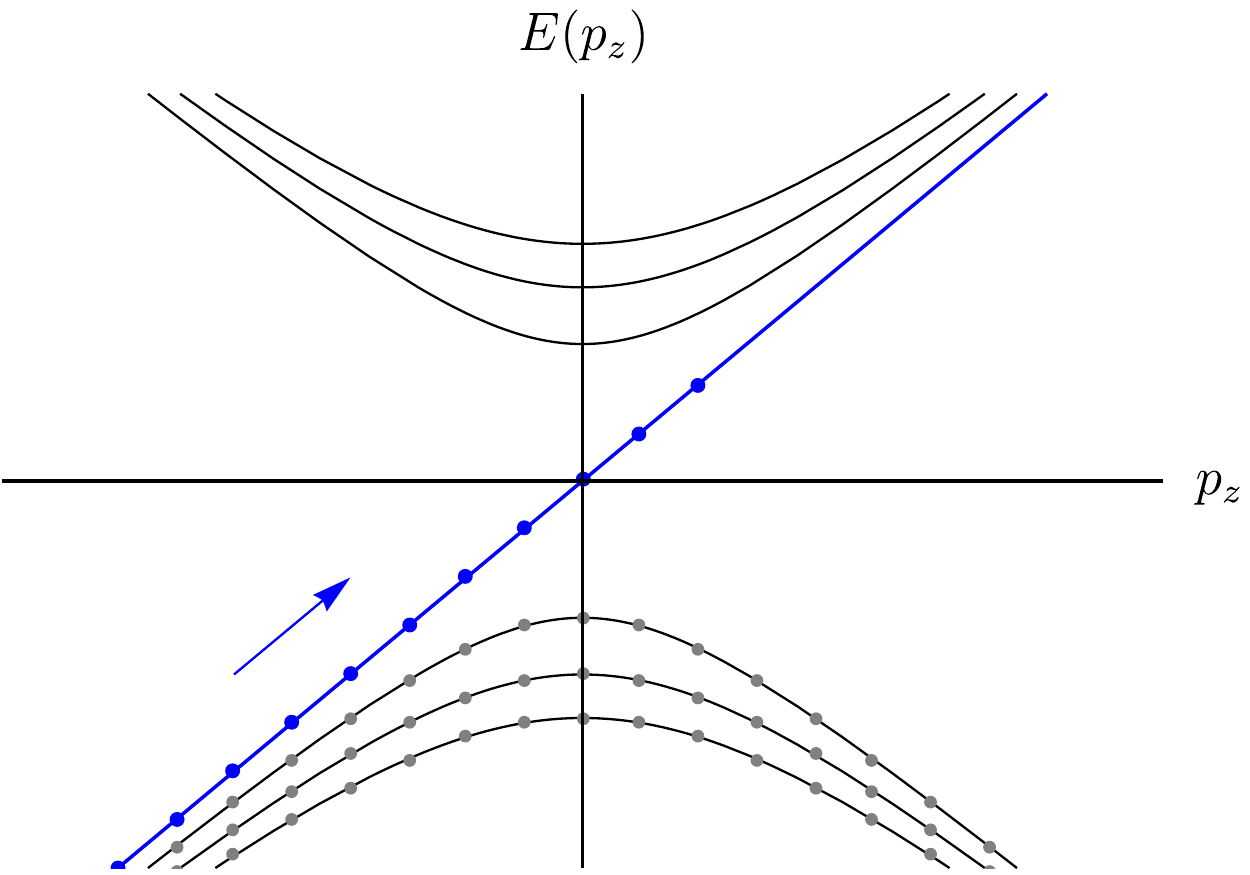}
        \caption{Spectral flow of parallel \textbf{B5, E} with the same conventions.}
        \label{fig:B5E}
             \includegraphics[width=220pt]{Kuvaajat/EM_kaikki_RH.pdf}\includegraphics[width=220pt]{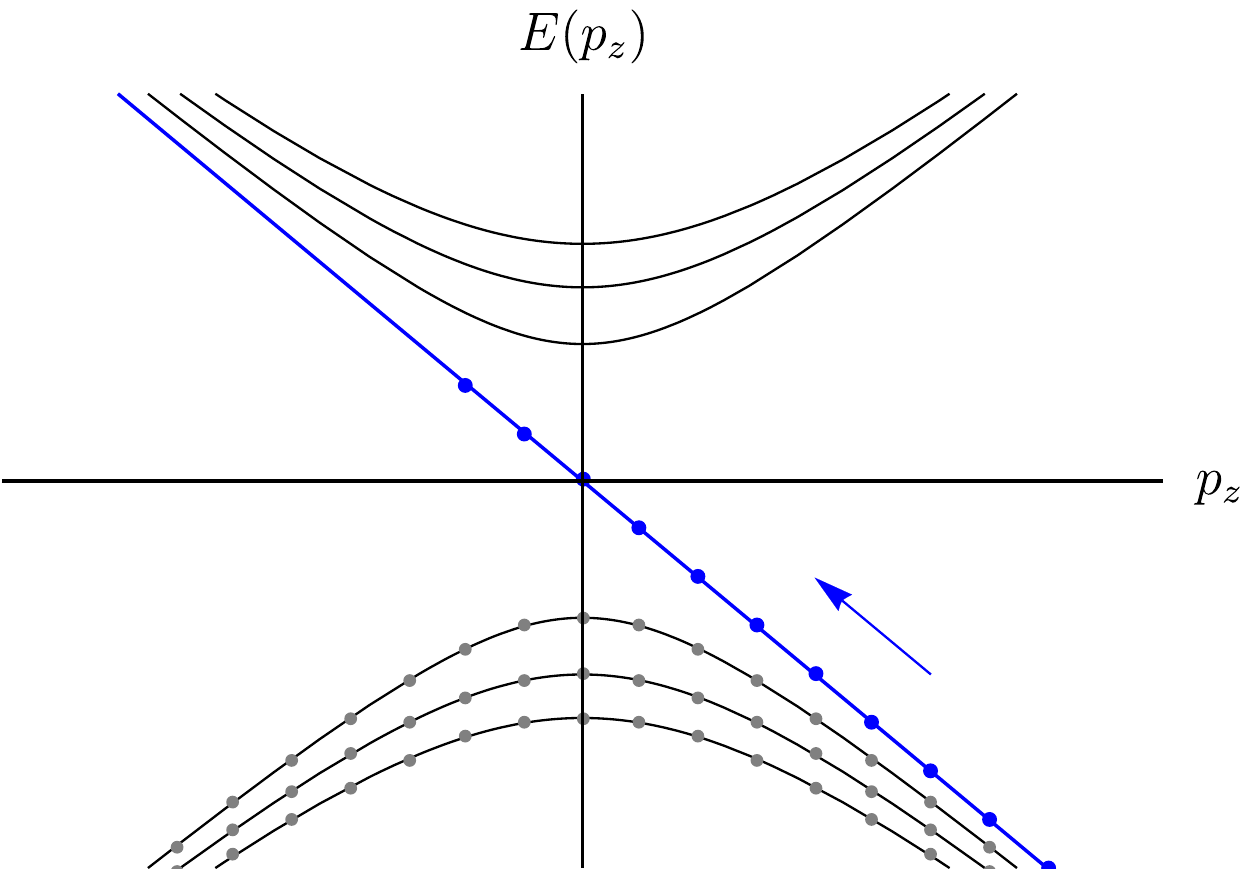}
        \caption{Spectral flow of parallel \textbf{B, E5} with the same conventions.}
        \label{fig:BE5}
        \includegraphics[width=220pt]{Kuvaajat/EM_kaikki_RH.pdf}\includegraphics[width=220pt]{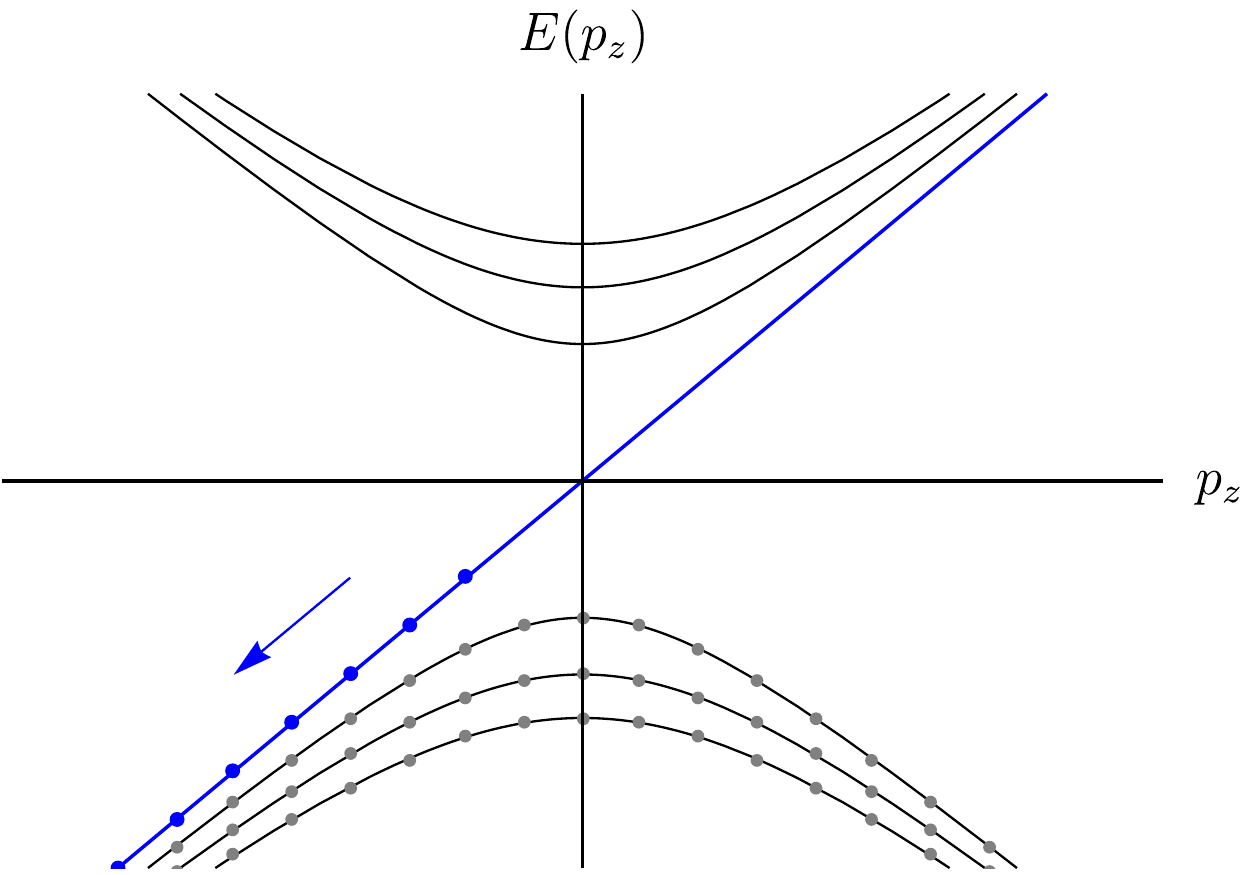}
        \caption{Spectral flow of parallel \textbf{B5, E5} with the same conventions.
        }
        \label{fig:B5E5}
       \end{figure*}
    \end{center}

Left and right chiral fermions may also couple independently to different gauge fields $A^+$ and $A^-$ depending on the chirality:
\begin{align}
\label{eq:axialham}
    H_\ch = \ch\sigma^i(p_i-qA^\pm_i)
\end{align}
This is often the case in condensed matter systems with pseudo gauge fields. We then define the axial vector potential \(A_5 = \half(A^+-A^-)\) with corresponding axial electric and magnetic fields \(\vek{B}^{5}\) and \(\vek{E}^{5}\), while the total vector potential is \(A = \half (A^+ + A^-)\). The corresponding currents are from
\eqref{eq:chiralcurrent}
\begin{align}
    \dot{j^0} & = \part j_+^0+\part j_-^0 = \frac{1}{2\pi^2} (E_z B^5_z + E^5_z B_z) \nonumber \\ 
   &=  \frac{-1}{8\pi^2}\levic F_{\mu\nu}F_{\rho\sigma}^5\\
     \dot{j_5^0} & = \part j_+^0-\part j_-^0 = \frac{1}{2\pi^2} (E_z B_z + E^5_z B^5_z)  \nonumber\\
     &=  \frac{-1}{16\pi^2}\levic \left( F_{\mu\nu}F_{\rho\sigma} + F_{\mu\nu}^5F_{\rho\sigma}^5\right).
\end{align}
This is the covariant chiral anomaly \eqref{eq:U1_anomaly_eqs}, represented as spectral flow under parallel electric and magnetic fields. The pictorial version for these equations in form of the LL the spectral flow can be found in Figs. \ref{fig:BE} to \ref{fig:B5E5}.

\subsubsection*{Weyl node at finite momentum}
In condensed matter systems the Weyl nodes are displaced from $p = 0$. Let us fix the symmetry by setting the node at $p_z = p_W$. It is straight-forward to see that shifting the momentum as \(\Tilde{p}_i = p_i \pm \delta_i^3 p_W\) in the Hamiltonian
\begin{align}
\label{eq:displacedham}
    H_\pm = \pm \sigma^i (\Tilde{p}_i-qA_i^\pm),
\end{align}
simply shifts the spectrum by $\pm p_W$. 

An axial anomaly can then arise from the momentum space structure itself, corresponding to e.g. the anomalous quantum Hall effect \cite{ZyuzinBurkov12, Burkov20}, since one has to shift $\Psi$ by $\Psi \to e^{\mp\im \vek{p}_W\cdot \vek{x}}\Psi$ to obtain the low-energy Weyl excitation $\Psi = e^{\pm\im \vek{p}_W\cdot \vek{x}} \Psi_W$. While this is not important for the spectrum of U(1) states, and the axial anomaly with parallel field, it is crucial in the case of the protected Fermi arcs, anomalous quantum Hall effect, as well as the torsional anomaly with the tetrads. 

\end{document}